\documentclass[10pt,a4paper]{article}
\usepackage[utf8]{inputenc}
\usepackage{mathtools}
\usepackage{amssymb}
\usepackage{float}
\usepackage[font=small,labelfont=bf]{caption}
\usepackage{subcaption}
\usepackage{graphicx}
\usepackage[export]{adjustbox}% http://ctan.org/pkg/adjustbox
\usepackage{hyperref}
\usepackage[version=3,arrows=pgf]{mhchem}
\usepackage[textsize=tiny]{todonotes}
\usepackage{authblk}
\usepackage{natbib}
\usepackage{setspace}
\usepackage[margin=1.0in]{geometry}
\usepackage{ulem}
\usepackage{tikz}
\tikzset{
    oxidation/.style={thick,red!80!black},
    reduction/.style={thick,blue!80!black},
}

\title{Modelling spatial constraints and scaling effects of catalyst phase separation
on linear pathway kinetics}
\author[1,2,3]{Nino Lauber}
\author[4]{Ondrej Tichacek}
\author[1,5]{Krishnadev Narayanankutty}
\author[1,6]{Daniele De Martino\thanks{daniele.demartino@ehu.eus, kepa.ruiz-mirazo@ehu.eus}}
\newcommand\CoAuthorMark{\footnotemark[\arabic{footnote}]}
\author[1,3]{Kepa Ruiz-Mirazo\protect\CoAuthorMark}
\affil[1]{\small Biofisika Institute(CSIC,UPV/EHU), Leioa, Spain}
\affil[2]{Donostia International Physics Center, Donostia-San Sebastian, Spain}
\affil[3]{ Department of Philosophy, University of the Basque Country, Donostia-San Sebastian, Spain}
\affil[4]{Institute of Organic Chemistry and Biochemistry of the Czech Academy of Sciences, Prague, Czech Republic}
\affil[5]{Department of Molecular Biology and Biochemistry, University of the Basque Country, Leioa, Spain}
\affil[6]{Ikerbasque Foundation, Bilbao, Spain}
\begin{document}

    \maketitle

    \begin{abstract}
Chemical reactions are usually studied under the assumption that both substrates and catalysts are well mixed (WM) throughout the system. Although this is often applicable to test-tube experimental conditions, it is not realistic in cellular environments, where biomolecules can undergo liquid-liquid phase separation (LLPS) and form condensates, leading to important functional outcomes, including the modulation of catalytic action. Similar processes may also play a role in protocellular systems, like primitive coacervates, or in membrane-assisted prebiotic pathways. Here we explore whether the de-mixing of catalysts could lead to the formation of micro-environments that influence the kinetics of a linear (multi-step) reaction pathway, as compared to a WM system. We implemented a general lattice model to simulate LLPS of an ensemble of different catalysts and extended it to include diffusion and a sequence of reactions of small substrates. We carried out a quantitative analysis of how the phase separation of the catalysts affects reaction times depending on the affinity between substrates and catalysts, the length of the reaction pathway, the system size, and the degree of homogeneity of the condensate. A key aspect underlying the differences reported between the two scenarios is that the scale invariance observed in the WM system is broken by condensation processes. The main theoretical implications of our results for ‘mean-field chemistry’ are drawn, extending the mass action kinetics scheme to include substrate initial “hitting times” to reach the catalysts condensate. We finally test this approach by considering open non-linear conditions, where we successfully predict, through microscopic simulations, that phase separation inhibits chemical oscillatory behaviour, providing a possible explanation for the marginal role that this complex dynamic behaviour plays in real metabolisms.

    	\medskip
		\noindent \textbf{Keywords}: Liquid-liquid phase separation (LLPS), biomolecular condensates, catalysed reaction kinetics, linear pathway, spatial constraint, lattice models, protocell chemistry
    \end{abstract}

    \section{Introduction}
        The interplay between physics and chemistry is key to reach a deep and comprehensive understanding of various aspects of cell physiology. 
        All metabolic processes take place under heterogeneous conditions and complex spatial constraints, often involving lipid bilayers and transport/transduction mechanisms supported by membrane compartments.
        Thus, the modelling of metabolism (or specific metabolic pathways) as a set of chemical reactions in open and well-mixed (WM) aqueous solution conditions is, at best, a crude first-approximation.
        This is becoming increasingly apparent in the last decades, even without taking into account lipid compartments, given the empirical evidence accumulating about the effects of `macromolecular crowding' and `biomolecular condensates' (membraneless compartments induced by liquid-liquid phase-separation) on various cellular functions \citep{Zhou:2008,Mitrea:2016,Banani:2017,Boeynaems:2018}.
        As a result, among other important aspects, a more realistic and precise characterization of protein activities \textit{in vivo} is being achieved.
        
        A particularly relevant derivative of the previous studies has to do with enzyme catalysis and the re-evaluation of classical biochemistry models based on `mass action kinetics' (MAK) assumptions ---e.g., the Michaelis-Menten equation--- which are not realistic under physiological conditions.
        Several extensions or alternatives to those traditional approaches have been proposed over the years.
        At a phenomenological level, we can mention `power-law' approximations to non-ideal behaviour (taking Savageau's seminal work as the main reference \citep{Savageau:1976,Savageau:1992}) or `fractal kinetics' (following Kopelman's works \citep{kopelman1986rate, kopelman1988fractal}).
        More recent contributions include `geometry-controlled kinetics' \citep{Benichou:2010,Benichou:2014} or `LLPS-modulated enzyme kinetics'\citep{OFlynn:2021,Peeples:2021}.
        In contrast with those deterministic approaches, more rigorous `stochastic kinetics' formalisms have also thrived, especially after Gillespie simulation algorithms \citep{Gillespie:1976,Gillespie:1977,Gillespie:1992} were introduced to solve the master equation for Markovian probability densities.
        However, the latter methods (which are much more accurate when molecular copy numbers are small) rely on the homogeneity of the reaction medium, which is not tenable under cytoplasmic spatial constraints.     

        In this context, different strategies can be tried (e.g. \citep{Mavelli:2010}) but lattice models constitute a well suited option, because they offer an explicit spatial framework to deal with phenomena like crowding or phase separation, and they can be combined with a microscopic, statistically coherent formulation of the chemical processes to be explored.
        For instance, Schnell and Turner \citep{Schnell:2004} (following Berry \citep{Berry:2002}), used this type of Monte Carlo simulations in a 2D lattice to study how mean-field `fractal kinetics' and `power-law' extensions should be amended when there is molecular crowding (i.e., obstacles to free diffusion) in the medium.
        One can find abundant lattice models applied to investigate liquid-liquid phase separation processes (one of our main targets of analysis here), but not so many have addressed their coupling with chemistry.
        Among the few that did, let us mention Glotzer \citep{Glotzer:1994}, who analysed phase separation development involving chemically active molecules (i.e., when the `scaffolds', the phase-separating species, are themselves reactants/products of some chemical transformation).\footnote{On similar lines, though focusing on the Michaelis-Menten (fully irreversible) scheme for enzymatic kinetics, Zhdanov \citep{zhdanov2000simulation} explored the possibility that the final product of a catalysed reaction undergoes phase separation, to conclude that the time scale to reach steady-state in the reactive dynamics gets much longer (if compared to the well mixed, non-segregated case).}
        This type of phenomenon (recently reviewed by Zwicker \citep{Zwicker:2022}) constitutes an open and promising area or research.

        The central motivation of this work is to investigate how a group of functional biomolecules may spatially organize themselves, thanks to weak associative interactions, to perform a collective chemical task (namely, multi-step catalysis on a linear reaction pathway).
        Therefore, we are interested in the analysis of LLPS processes in which the biopolymers are both the `clients' and `scaffolds' of the condensate, to test in particular how that modifies their catalytic action.
        The standard way of addressing these issues experimentally, in well-controlled in vitro conditions, tends to split up those two roles, making use of some biomolecules (e.g.: PEG and dextran \citep{Davis:2015}, or polySH3 and polyPRM \citep{Peeples:2021}) to induce the condensates, and then analysing the effects of phase separation on others performing catalytic function.
        Yet this is just a practical, methodological commodity.
        Nothing precludes the catalytic biopolymers from generating the condensate themselves under in vivo conditions.
        Quite the contrary: enzymes with an inherent oligomerizing propensity often display multivalent interactions, which give them the potential to trigger LLPS processes.
        As a matter of fact, a remarkable number of metabolism-related enzymes have been observed to form condensates in bacteria, yeast and other organisms -- as it is more thoroughly reviewed in \citep{Prouteau:2018}.

        Our approach here is quite general, not restricted to water-soluble enzymes/catalysts.
        Very important physiological and metabolic processes (think of phospholipid synthesis, like the Kennedy pathway, signalling pathways or electron-transfer chains, just to mention a few examples) take place around or within membranes.
        Therefore protein aggregation-disaggregation dynamics in the context of 2D lipid bilayers may play a key role in complex cell behaviour.
        Furthermore, under much simpler conditions, the combination of chemical transformations and physical/spatial constraints also seems to be critical for the process of emergence of biological cells (hence the current weight of the `protocell camp' within origins-of-life research: see, e.g., \citep{Chen:2010,ShirtEdiss:2017,Toparlak:2019,Gozen:2022}).
        Phase separation into `coacervates' \citep{BungenbergDeJong:1929} was actually long ago proposed by Oparin \citep{Oparin:1938} as a crucial step for prebiotic chemistry to unfold.
        The hypothesis was left aside for many years (in particular, with the advent of liposome science \citep{Bangham:1974}, favouring vesicles as protocell models \citep{Morowitz:1988,Luisi:2006}; \citep{Hanczyc:2008}) but has been revisited and gained new momentum in recent years \citep{Donau:2020}.
        Taking all these considerations into account, we will formulate the chemical part of the problem here in rather general terms, without sticking to the standard mechanism of catalysis (as it is performed by enzymes in open aqueous solution -- i.e., the Michaelis-Menten scheme), but with a broader view of catalytic action that could apply to allegedly prebiotic or membrane-associated linear reaction pathways.

        The article is structured as follows.
        First, in Sect.~\ref{sec:model}, we briefly describe the lattice model employed to explore microscopically the LLPS dynamics of a group of catalysts
        \citep{Sear:2003, Jacobs:2017, Jacobs:2021, carugno2022instabilities, shrinivas2021phase}
        %(building upon Kawasaki's formulation of the original Ising model \citep{Ising:1925,Kawasaki:1966}), 
        which is extended to simulate a simple chemistry (a catalysed, multi-step, linear reaction pathway) by considering the various reactants as `random walkers' on the lattice.
        The results of our theoretical work are then reported in Sect.~\ref{sec:results}, divided in a first subsection~\ref{sec:res_main} where the main outcomes of the simulation runs are summarized (in terms of the effects that different equilibrium configurations of the catalysts have on the global reaction time required to complete the pathway) and a second one \ref{sec:res_thry} in which the implications for `mass action kinetics' are drawn, and the model is also tried for a more complex chemistry taking place in non-equilibrium conditions.
        We conclude with a short discussion about the potential and limitations of our approach, indicating possible lines of work for the future.

    \section{The model}
    \label{sec:model}
\begin{figure}[ht]
        	\centering
        	\begin{subfigure}[b]{0.5\textwidth}
        		\includegraphics[width=\textwidth]{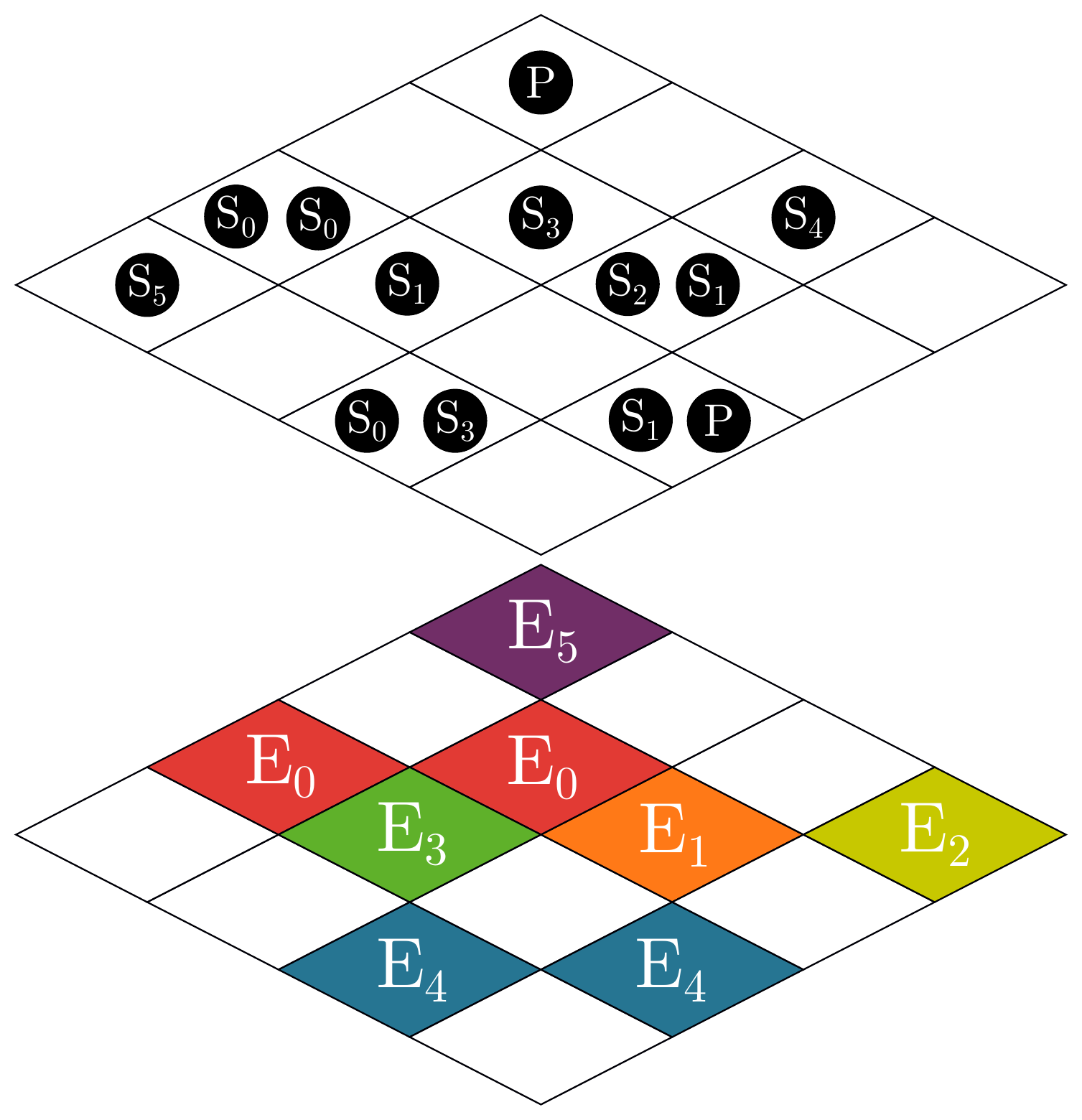}
        		\caption{}
        		\label{fig:latt_mod}
        	\end{subfigure}
        	\hspace{2ex}
        	\begin{subfigure}[b]{0.2\textwidth}
        		\includegraphics[width=\textwidth]{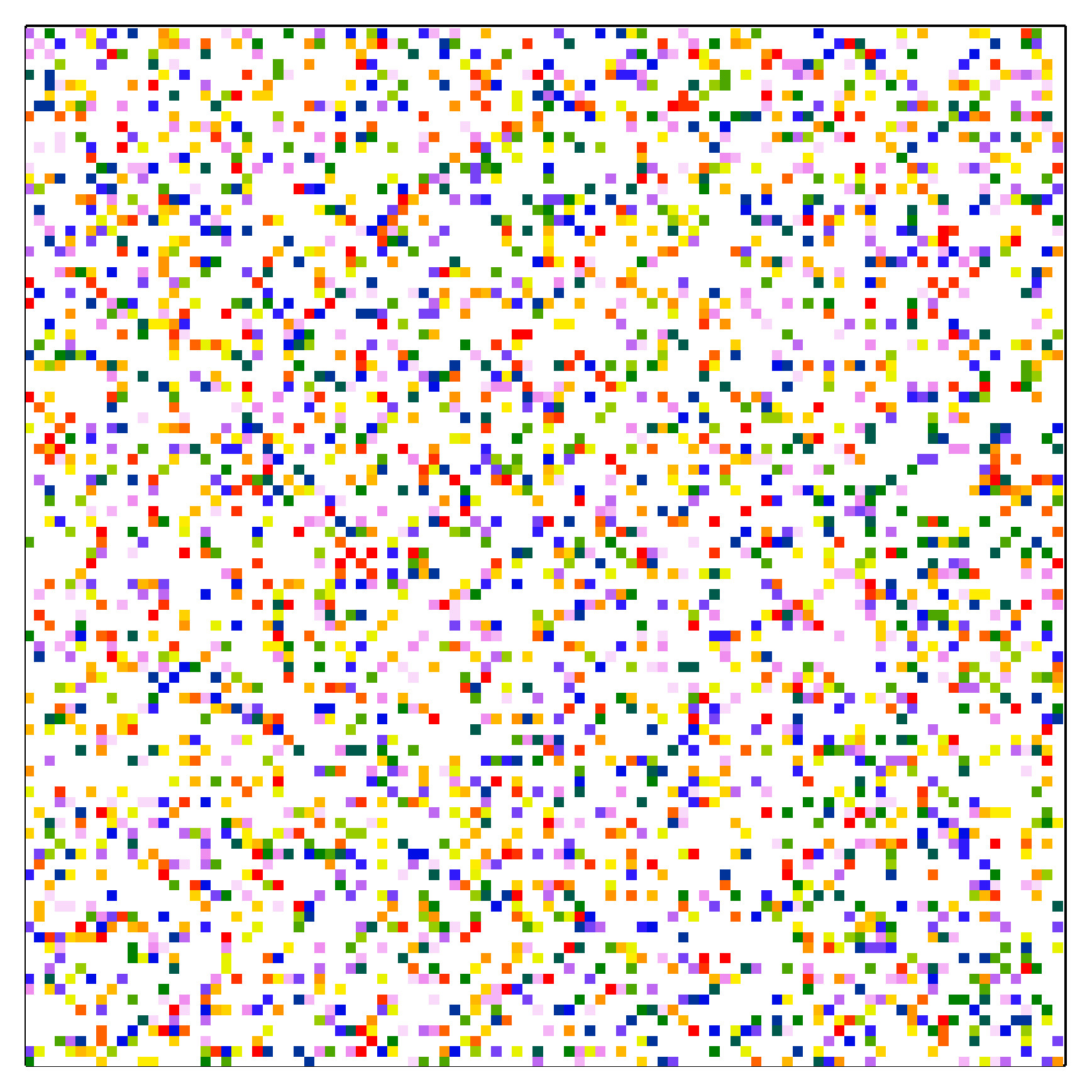}
        		\caption{}
        		\label{fig:wm_syst}
        		
        		\includegraphics[width=\textwidth]{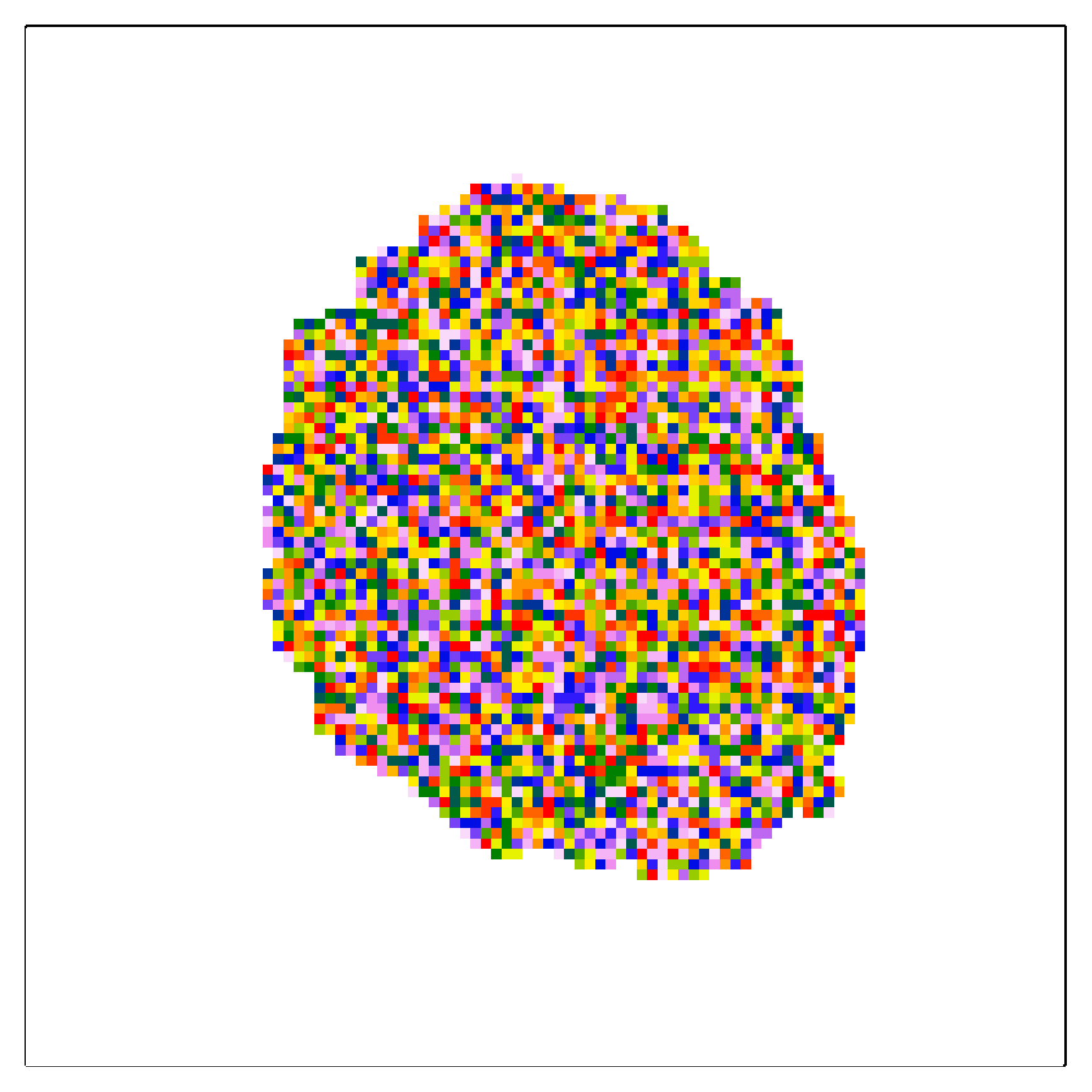}
        		\caption{}
        		\label{fig:ps_syst}
        	\end{subfigure}
        	\caption{
            \textbf{(a)} The lattice model. Biomolecules $E_i$ occupy single squares on the bottom grid (the catalyst-lattice),  where different colors stand for different catalysts (white squares represent the solvent).
            Multiple substrates $S_i$ can occupy a single square on the top grid (the substrate-lattice).
            \textbf{(b)} Typical equilibrium configuration from simulations with lattice size $d=100$, pathway length $L=20$, surface fraction $\phi_{tot}=0.3$, interaction strength $J=1.0$ and inverse temperature  $\beta=0.01$.
            \textbf{(c)} Typical equilibrium configuration for the same parameters as in \textbf{(b)}, except for $\beta=5.0$
            }
        	\label{fig:mult_latt}
        \end{figure}

        We consider a linear reaction pathway where an initial substrate $S_{0}$ is transformed through consecutive steps and intermediary substrates $S_{\mu}$ into a final product $P$.
        Each reaction step is catalysed by a specific biomolecule\footnote{This could be an enzyme/protein, as it is usually the case in extant metabolic pathways, but also some other organocatalyst, like an oligopeptide or an RNA-oligonucleotide, in the context of a protocell or under prebiotic conditions.} $E_{\mu}$, \ce{S_{\mu-1} ->[E_{\mu}] S_{\mu}}, so we have:
        \begin{equation}
            \ce{S_{0} ->[E_{1}] S_{1} ->[E_{2}] S_{2} \dots S_{L-1} ->[E_{L}] P }
            \label{eq:cascade}
        \end{equation}
        where we denote the number of catalysts $L$, corresponding to the length of the reaction pathway.
        It is assumed that these $L$ catalysts have a mild attraction to each other (typically, through multivalent interactions) and therefore they can undergo phase-separation.
        As depicted in Fig.~\ref{fig:mult_latt}, in order to simulate the LLPS of the catalysts, space is discretized  into a two dimensional, $d\times d$ regular lattice, where each site-$i$ is in a state $\sigma_i=0,1\dots L$ representing the type of catalyst inhabiting the respective site, and $0$ representing the solvent.
        The energy of a certain configuration of catalysts on the lattice is given by:
        \begin{equation}
            H (\vec{\sigma}) = -\sum_{\langle i,j\rangle}J(\sigma_i,\sigma_j)
            \label{eq:E_latt}
        \end{equation}
        The function $J(\sigma,\sigma')$ determines the strength of interaction between two compounds and takes the form of a symmetric matrix $J_{\sigma\sigma'}=J_{\sigma'\sigma}$.
        The brackets in the sum stand for neighbouring lattice sites. 
        In this work we do not take into account solvent-solvent nor solvent-catalyst interactions: i.e., $J_{00}=J_{\sigma 0}=0$.
        Notice that in our convention $J_{\sigma\sigma'}>0$ corresponds to attraction, while $J_{\sigma\sigma'}<0$ corresponds to repulsion.
        Following standard statistical mechanics, we consider equilibrium configurations whose probability is given by the Boltzmann-Gibbs distribution
        \begin{equation}
            P(\vec{ \sigma}) \propto e^{-\beta H (\vec{\sigma})},     
        \end{equation}
        where $\beta=1/T$ is the inverse temperature.
        We limit ourselves to attractive interactions and for further simplicity, we assume that all catalysts attract each other with the same strength: i.e., $J_{\sigma\sigma'}=J,\ \forall \sigma,\sigma'\neq 0$.
        This last assumption will be slightly modified when discussing the problem of the homogeneity of the condensate, later on.
        Our approach builds upon previous work on the thermodynamics of phase separation following these premises: \citep{Sear:2003, Jacobs:2017, Jacobs:2021, carugno2022instabilities, shrinivas2021phase,lauber2022statistical}.
        
        Each catalyst $E_\mu$ is present on the lattice in a certain volume or, rather, surface fraction $\phi_\mu$ -- with $\phi_{tot}=\sum_{\mu}\phi_\mu$ being the total surface fraction of catalysts on the lattice.
        We further assume that each catalyst has the same surface fraction, i.e. $\phi_\mu=\frac{\phi_{tot}}{L}$.
        In that case, it is easy to show that this model is formally equivalent to the Ising model at conserved or constant magnetization, which is known to undergo phase separation.
        This behaviour can be captured qualitatively by the regular solution model and easily simulated in the lattice with a Metropolis Monte Carlo algorithm (more precisely, implementing the Kawasaki rule \citep{Kawasaki:1966} for the conservation of enzyme copy numbers).
        The diffusive dynamics of the substrates is modelled as a standard `Brownian motion' phenomenon: i.e., as a simple lattice random walk.
        We assume perfect catalysis and irreversible reactions: every time an intermediate substrate encounters its corresponding catalyst in the pathway, it gets immediately transformed into the next intermediate substrate in the linear chain, which starts its random walk from the position of the catalyst that has led to it.
        The model and its overall features are illustrated in Fig.~\ref{fig:mult_latt}. 
    
        Our aim in the current paper is to investigate, within this simple setting, how the behaviour of the reaction pathway, defined by Eq.~\ref{eq:cascade}, is affected if the catalysts are well-mixed (as shown in Fig.~\ref{fig:wm_syst}) or if they phase separate into a surface droplet (as shown in Fig.~\ref{fig:ps_syst}).
        In particular, in order to quantify the behaviour of the reaction pathway we will compute the mean reaction-time $\overline{t}_{R}$, which is the average time for an initial substrate $S_{0}$ to complete the reaction; namely, to be transformed into the final the product $P$.
        Therefore, in the context of our simulations, $\overline{t}_R$  corresponds to the average time it takes a substrate that is initialized as $S_0$ to physically encounter all the catalysts $E_\mu$ in the correct order, as defined by the reaction pathway in Eq.~\ref{eq:cascade}.
        Depending on the equilibrium configuration of the condensate (and on the other parameters of the simulations), a number of different outcomes were statistically collected, as we report next.

    \section{Results}
    \label{sec:results}
        \subsection{Main simulation results}
        \label{sec:res_main}
        \subsubsection{Substrate-catalyst affinity accelerates pathway completion in PS systems.}
        \label{sec:res_main1}
            As the first result, let us show how a minimal degree of affinity between the substrates and the catalysts is critical for the condensates to enhance catalysis through a concentration mechanism.
            With this aim, we introduce parameter $I$, which quantifies the strength of the substrate-catalyst interaction\footnote{As with catalyst-catalyst interactions, we could in principle introduce a matrix encoding for  different interaction strengths among different substrates and catalysts, but here, for sake of simplicity, we just consider homogeneous interactions.}.
            The presence of such interaction amounts to a bias in the random walk simulating the substrate dynamics.
            Under these conditions, then, the rate at which the substrate leaves regions occupied by solvents and enters regions occupied by the catalysts is simply modulated by a Boltzmann weight:
            \begin{equation}
                W(i \to j | \sigma_i=0, \sigma_j>0) \propto e^{\beta I}  
            \end{equation}
            and inversely for the rate $W(i \to j | \sigma_i>0, \sigma_j=0)$.
            \begin{figure}[ht]
        		\centering
        		\includegraphics[width=0.7\textwidth]{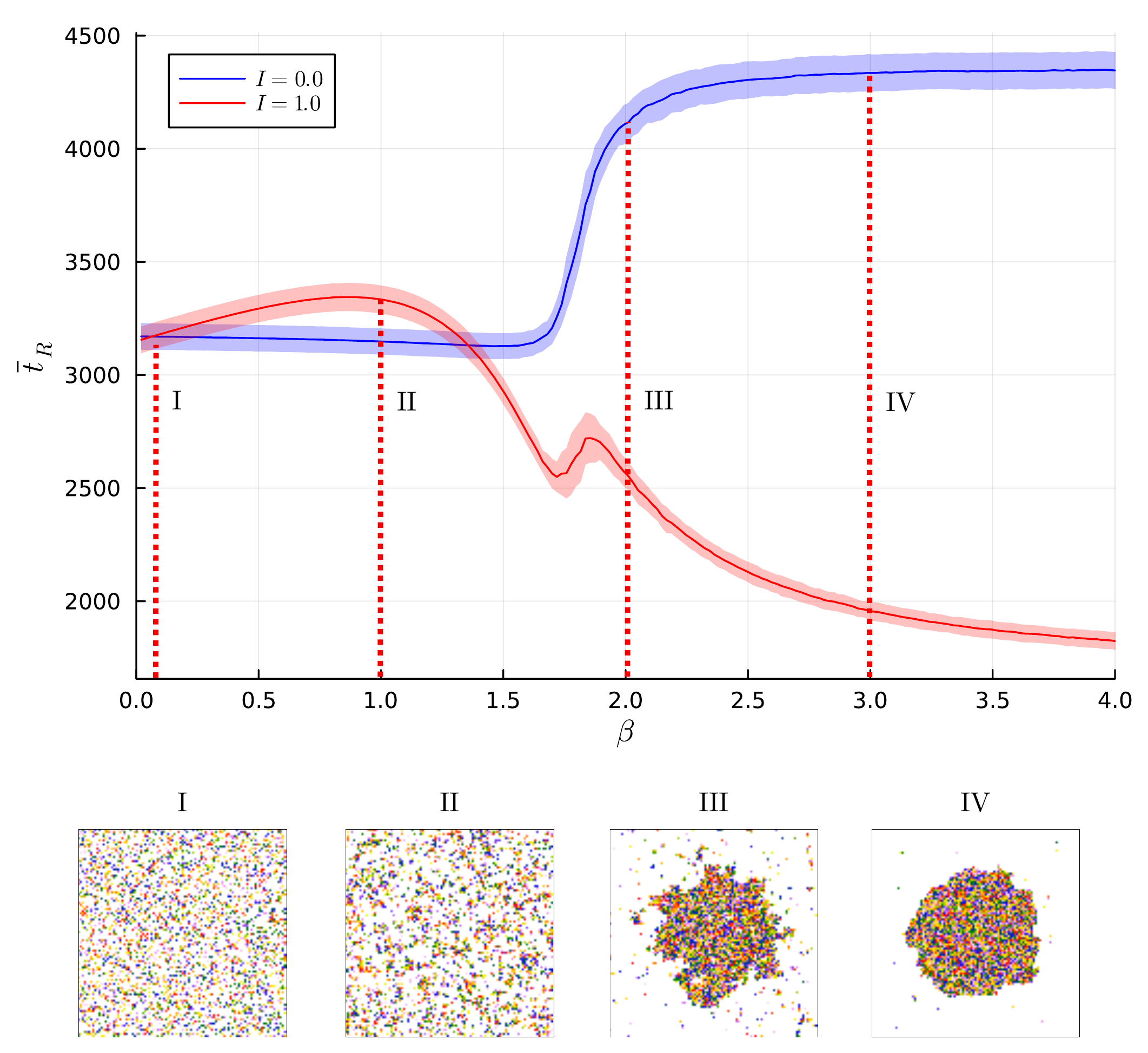}
        		\caption{
                Average reaction time $\overline{t}_R$ as a function of the inverse temperature $\beta$ from simulations performed at two different levels of substrate-catalyst interaction: $I=0.0$ (no attraction, blue curve) and $I=1.0$ (attraction, red curve). Otherwise, parameters were fixed (lattice size $d=100$, pathway length $L=20$, catalyst surface fraction $\phi_{tot}=0.3$, catalyst-catalyst interaction strength $J=1.0$). The curves have been averaged over 72 independent simulations with the error bar showing the standard deviation.
                Boxes I-IV: snapshots of  equilibrium configurations  at $\beta=0.0,1.0,2.0,3.0$.
                }
        	   \label{fig:tRdepI}
            \end{figure}
            In order to investigate the effect of the substrate-catalyst interaction $I$ on the reaction time through the spectrum of systems ranging from well mixed to fully phase-separated,
            we computed the mean time to react $\overline{t}_R$  from numerical simulations at increasing $\beta$, that is, transitioning from a WM-system to a PS-system.
            In Fig.~\ref{fig:tRdepI} we illustrate graphically this effect by comparing two systems: one with no substrate-catalyst interaction and one with mild attraction between substrates and catalysts. 
            In the former case $\overline{t}_R$ is constant for low $\beta$ but starts to increase once a condensate starts to form (region III in Fig.\ref{fig:tRdepI}).
            In the case of substrate-catalyst attraction, $\overline{t}_R$ shows an initial increase, as small droplets start to form (region II in Fig.\ref{fig:tRdepI}) but then decreases, once the droplets percolate and form a macroscopic condensate.
            In general $I>0$ means then that the catalysts can ``trap'' the substrate molecules within the condensate, where they can find the right enzyme in the pathway much quicker (in comparison to the WM-case), given their higher local concentration.
            This mechanism also explains the uptick in region II, because there are droplets that can trap the substrate, but evidently lack exemplars of all the catalysts of the pathway and thus lead to higher concentrations of incomplete intermediates.
            In contrast, for no attraction at all ($I=0$), in the PS regime the condensate is not able to keep the substrate inside, with intermediates diffusing in and out and finally leading to less efficient catalysis, compared to the WM regime.
            A certain degree of attraction is therefore necessary to enhance catalysis in the PS regime, so in the following sections we will assume a mild substrate-catalyst attraction (i.e., the condensate will be considered, by default, a ``trapping'' condensate).

        \subsubsection{Multi-step reaction kinetics is not scale invariant in PS systems: pathway length and size effects}
        \label{sec:res_main2}
            Next we turn our focus on the dependence of the average reaction time $\overline{t}_R$ on the pathway length $L$.
            The first important difference between the WM and PS cases concerns the system  size $d$: a WM system is scale invariant while a PS is not, since the time required for a substrate $S_0$ to reach the condensate increases with the system size.
            In the appendix~\ref{sec:tR_wm} we derive the following formula for the dependence of $\overline{t}_R$ on $L$ and $\phi_{tot}$ for a WM system ($a$ is a constant with regard to $L$ and $\phi_{tot}$, but dependent on several other parameters, like the lattice geometry, the diffusivity of the substrate, etc.):
            \begin{equation}
                \overline{t}_R^{\ wm} = a\ \frac{L^2}{\phi_{tot}}
                \label{eq:tRwm}
            \end{equation}
            In essence this time is inversely proportional to  the probability of reaching the correct enzyme by the substrate during its random walk.
            Now, for the PS case one can assume that, once inside the condensate, and if the latter is homogeneous, the substrate will experience an environment  similar to the WM case but  with a local volume fraction of enzymes  of $\phi'_{tot}\sim 1$.
            However, as already mentioned above, the initial substrate $S_0$ first needs to find the droplet.
            Thus the mean reaction time is delayed by the initial time to ``hit'' the droplet, which we will denote as $t_H$ (the hitting-time).
            Thus, the following expression is derived:
            \begin{equation}
                \overline{t}_R^{\ ps} = aL^2 + t_{H}(\phi_{tot},d)
                \label{eq:tRps}
            \end{equation}
            In other words, $t_{H}(\phi_{tot},d)$ implies an offset for the mean reaction time that depends on $\phi_{tot}$ and the lattice size $d$.
            In the appendix~\ref{sec:tR_ps} we derive an exact formula for the case with circular symmetry, finding the scaling to be:
            \begin{equation}
                t_{H}(\phi_{tot},d) = b d^2 (1-\phi_{tot})g(\phi_{tot})
            \end{equation}
            
            We tested these predictions for the mean reaction time by performing simulations and computing  $\overline{t}_R$ for systems of variable pathway-length $L$, different lattice size $d$ (with a couple of fixed $\beta$ values, one leading to a WM system and the other to a PS system).
            \begin{figure}[ht]
                \centering
                \includegraphics[width=0.9\textwidth]{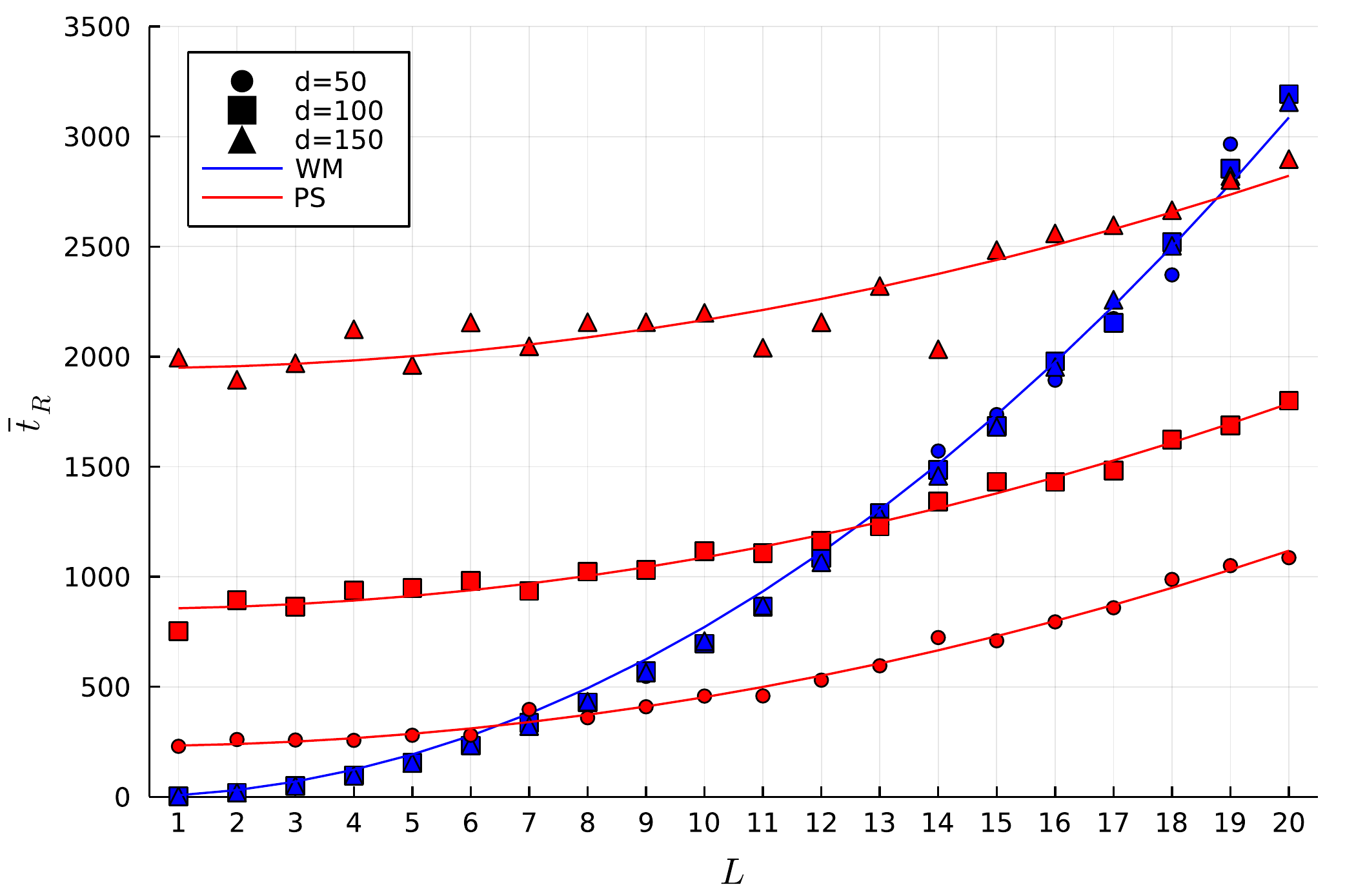}
                \caption{Average reaction time $\overline{t}_R$ as a function of the pathway length $L$ ($\phi_{tot}=0.3$, $J=1.0$). Simulations were carried out for different lattice sizes $d=50,100,150$ (indicate as circles, square and triangles respectively) and comparing WM-systems (blue, $\beta=0.0$ and $I=0.0$) with PS-systems (red, $\beta=5.0$ and $I=1.0$). Solid lines:  fit of the simulation data using Eq.~\ref{eq:tRwm} and Eq.~\ref{eq:tRps} for the WM- and the PS-systems, respectively.}
                \label{fig:tVn}
                \end{figure}
                
            Our results are illustrated in Fig.~\ref{fig:tVn}.
            The values for $\overline{t}_R$ obtained in the WM-case coincide for all choices of $d$ (scale invariance) and they follow a parabolic trend that is well fitted by Eq.~\ref{eq:tRwm}.
            By contrast, in the PS-case one can see that, while the $\overline{t}_R$ values obtained lie on parabolas with the same slope, there is an offset that increases with the system size, which is in good accordance with our prediction in Eq.~\ref{eq:tRps}, based on considering the size dependence of the hitting time.
            In general, we can state that there is a certain critical length $L_c$ of the reaction path above which the mean reaction time is shorter for a PS system (in comparison to a WM system) and thus catalysis is enhanced by the formation of a condensate.
            We can get a formula for $L_c$ by putting together Eq.~\ref{eq:tRwm} and Eq~\ref{eq:tRps}:
            \begin{equation}
                L_c = d\cdot\sqrt{\frac{b\,\phi_{tot}\,g(\phi_{tot})}{a}}
                \label{eq::Lcrit}
            \end{equation}
            Thus our equations predict that $L_c\propto d$ and this is verified by numerical simulations (as it is easy to infer from Fig.~\ref{fig:tVn}, where we obtain $L_c \sim 6,12,18$ for $d=50,100,150$ respectively).

        \subsubsection{Dependence of the reaction-time on the homogeneity of the condensate}
        \label{sec:res_main3}
            Up to this point all interactions were assumed to be homogeneous. 
            However, in a more realistic setting different enzymes/catalysts will show different modes and intensities of attraction between each other due to geometry, charge distribution, ionization, functional groups, etc.
            This leads to a variety of multivalent interactions that determine the structure and dynamical properties of the condensates which, in turn, affect their catalytic performance.
            In order to analyse this aspect, we introduce here a certain degree of heterogeneity by changing the catalyst-catalyst interaction $J$ back from a single scalar parameter to a real matrix, where the different entries $J_{\sigma \rho}$ reflect the strength of interaction between each pair of catalysts $\sigma,\rho$ (as we said above, this matrix will be symmetric: i.e., $J_{\sigma \rho}=J_{\rho \sigma}$).
            We consider a random matrix whose entries will be identical and independently distributed gaussian random variables with average value that we call $\mu_J$ and standard deviation that we call $\sigma_J$.
            We set  $\mu_J=1$ and restrict ourselves to the case $\sigma_J/\mu_J<<1$, in such a way that the majority of interactions are attractive and the system phase separates with a thermodynamic behavior that is slightly perturbed with respect to the homogeneous case ($\sigma_{J}=0$).
            Similar to what we did in Sec.~\ref{sec:res_main1}, the global average reaction time was computed through numerical simulations as a function of $\beta$ and for diverse values of $\sigma_J$.
            \begin{figure}[ht]
        		\centering
        		\begin{subfigure}[b]{0.75\textwidth}
                    \includegraphics[width=\textwidth]{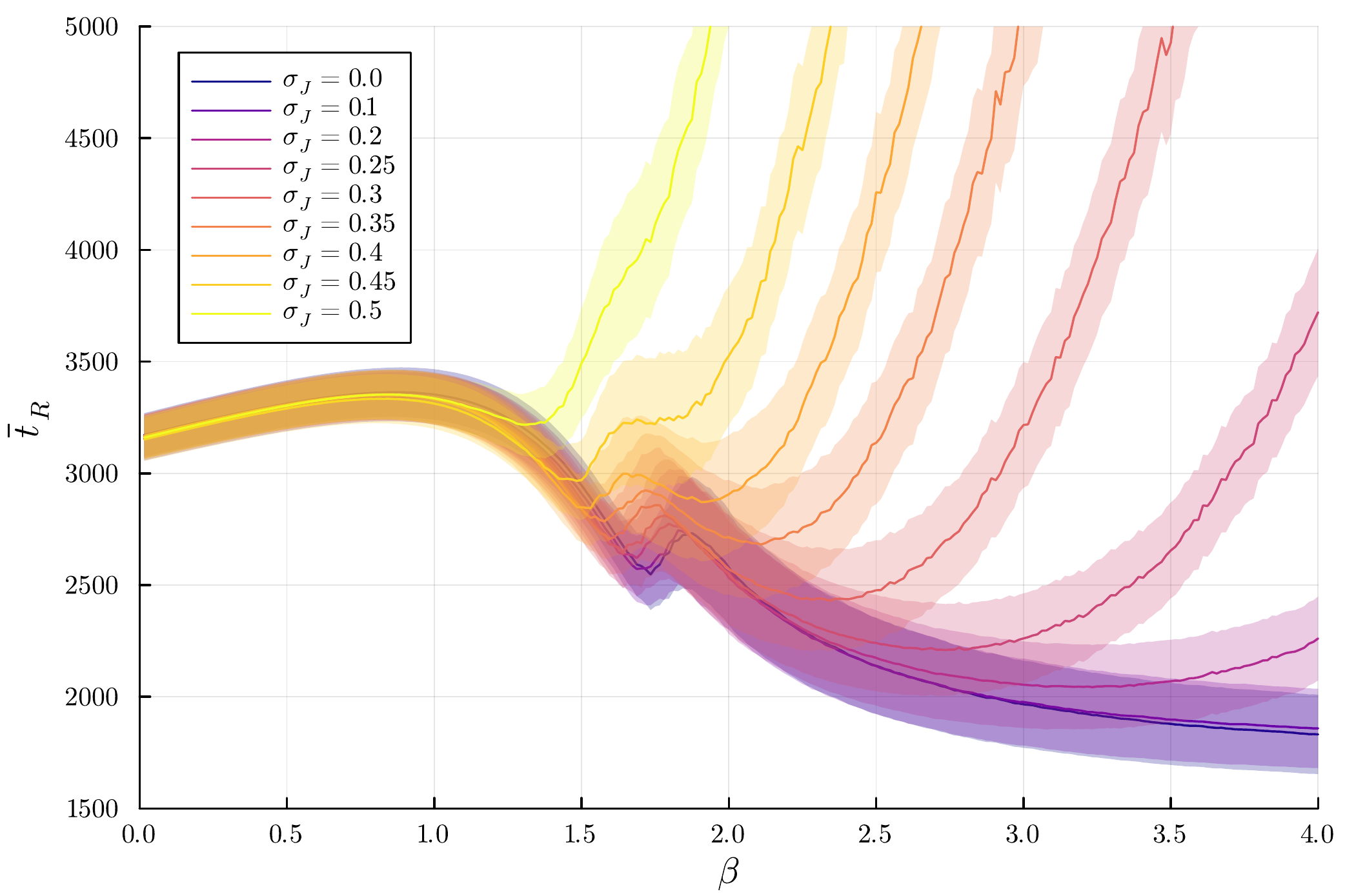}
        		    \caption{}
        		    \label{fig:tRVb}
        		\end{subfigure}
        		\begin{subfigure}[b]{0.2\textwidth}
        		    \includegraphics[width=\textwidth]{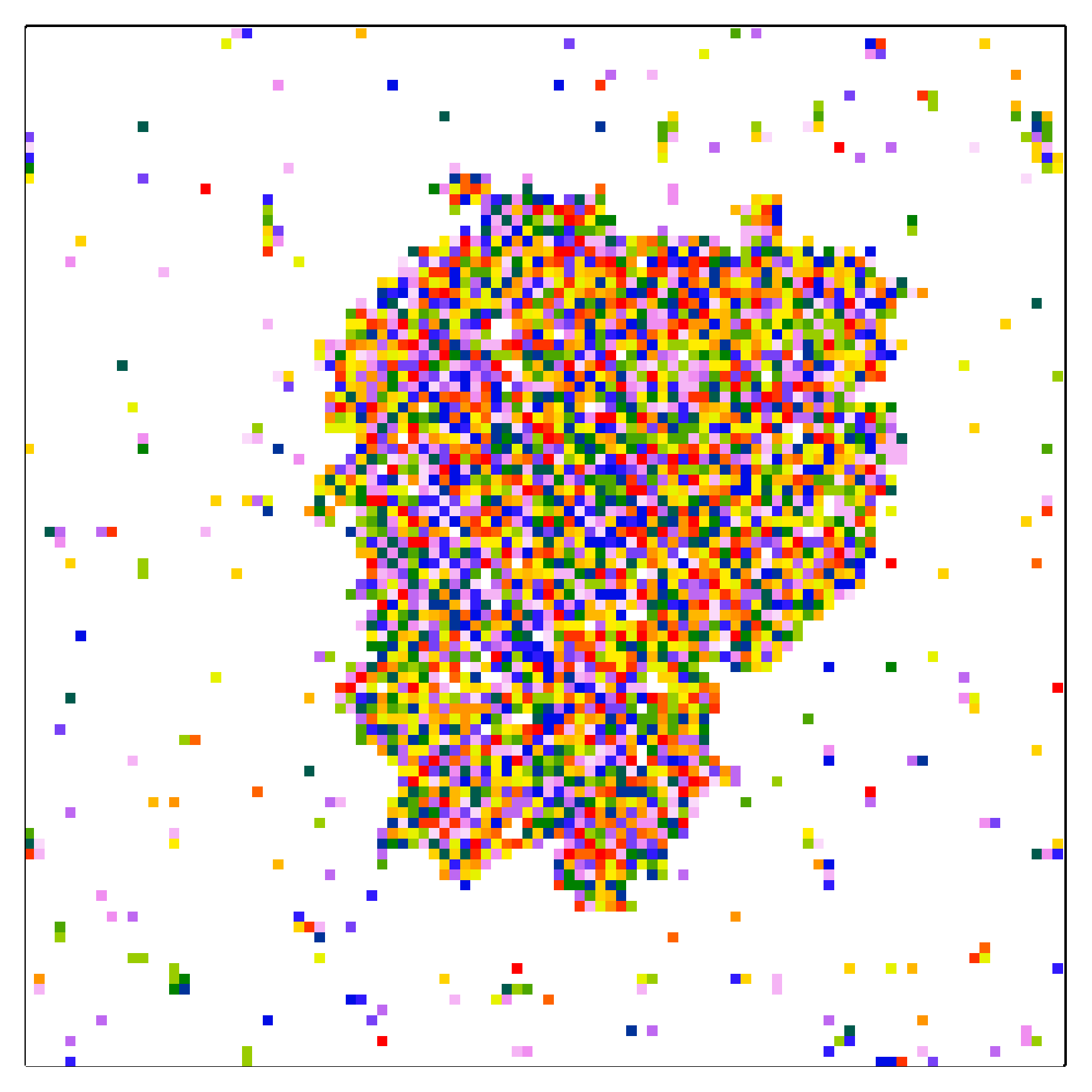}
                    
                    \includegraphics[width=\textwidth]{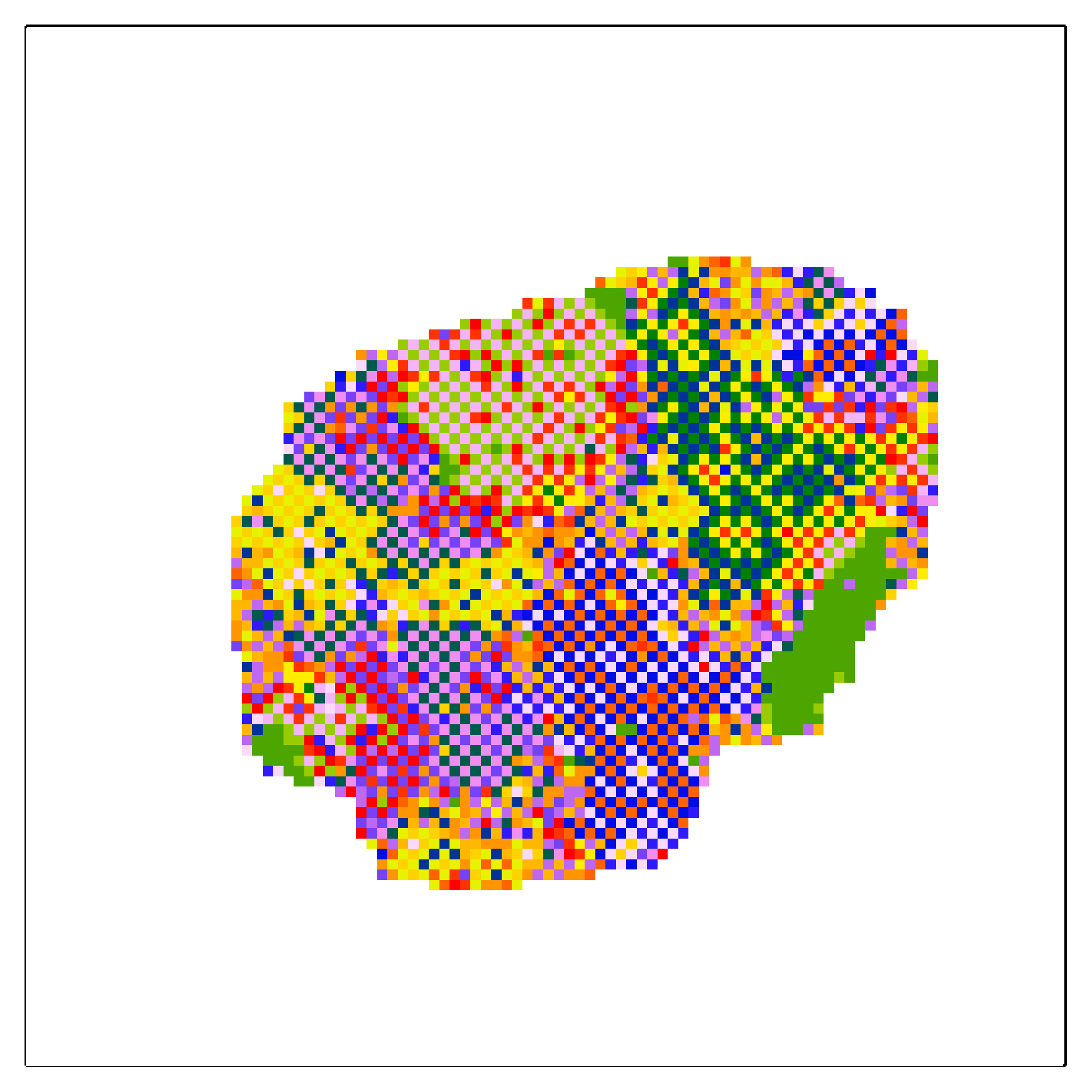}
        		    \vspace{1ex}
        		    \caption{}
        		    \label{fig:ps_inhom}
        		    \vspace{1ex}
        		\end{subfigure}
        		\caption{\textbf{(a)} Global average reaction time $\overline{t}_R$ as a function of the inverse temperature $\beta$ (lattice  size $d=100$, pathway length $L=20$, enzyme volume fraction $\phi_{tot}=0.3$,  mean catalyst-catalyst interaction strength  $\mu_J=1.0$, substrate-catalyst interaction strength $I=1.0$), for different $\sigma_J$ (different line colors) \textbf{(b)} Upper panel: equilibrium configuration  for a system with $\sigma_J=0.3$ and $\beta=2.0$. Lower panel: equilibrium configuration with $\sigma_J=0.3$ and $\beta=5.0$}
        		\label{fig:tRdepS}
        	\end{figure}

            Our results are reported in Fig.~\ref{fig:tRVb}.
            Similarly to the homogeneous case (Fig.~\ref{fig:tRdepI}), the mean reaction time initially increases when increasing $\beta$ (this corresponds to the small surface-droplet forming region) but then decreases, once $\beta$ is in the region where a macroscopic condensate starts to form.
            In contrast to the homogeneous case, however, where further increasing $\beta$ does not change $\overline{t}_R$, we can see that for larger values of $\sigma_J$ the mean reaction time starts to increase again.
            In other words, the trend is non-monotonous, with a well-defined minimum -- or an optimal inverse temperature $\beta_o$ -- where catalysis is more efficient.

            This non-monotonous behavior is mainly due to the fact that, for lower temperatures, the condensate itself is not well-mixed and substrates typically travel longer distances before encountering the right catalyst.
            This explanation is coherent with a correlation that we observe between the increase of reaction times and the formation of patches in the condensate for lower temperature (as illustrated in Fig.~\ref{fig:ps_inhom} upper panel vs  lower panel).
            The presence of these patches (for which there is evidence in the literature on biomolecular condensates \citep{Fare:2021}) makes it more difficult for a substrate inside the droplet to find the catalysts in the right order (larger $\sigma_J$ value means a more ``patchy'' droplet, as the interactions are more heterogeneous).

        \subsection{Theoretical implications}
        \label{sec:res_thry}
        \subsubsection{A mean-field approach to model reaction kinetics for  PS systems.}
        \label{sec:res_thry1}
\begin{figure}[ht]
        		\centering
        		\includegraphics[width=0.8\textwidth]{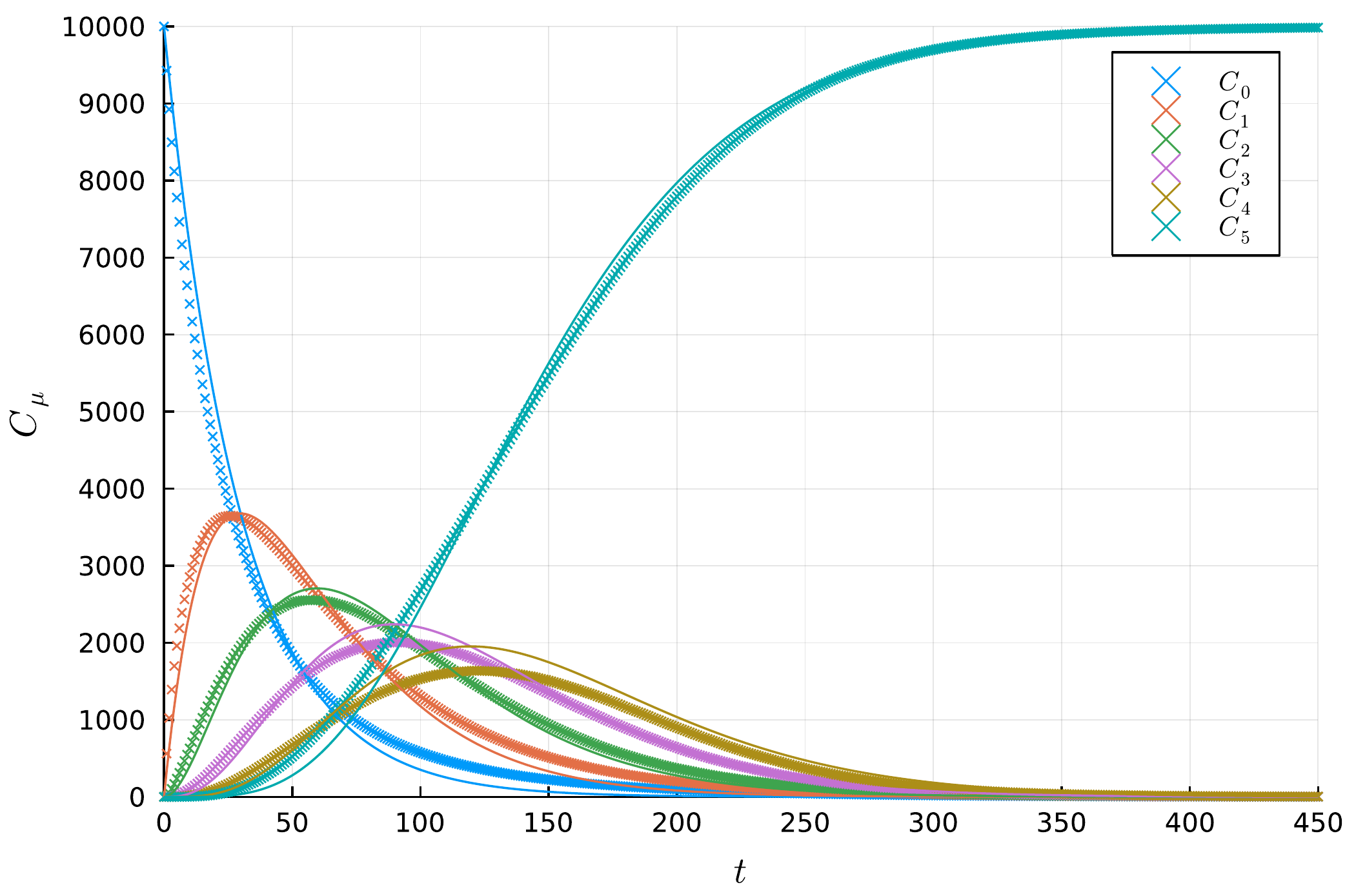}
                \caption{Time dynamics for the substrate concentrations $C_\mu(t)$ from full microscopic simulations (points) and the analytical solution of the mass actions ODE (lines) in a WM-system (lattice size $d=100$, pathway length $L=5$, catalyst surface fraction $\phi_{tot}=0.3$, catalyst-catalyst interaction strength  $J=1.0$, substrate-catalyst interaction strength $I=0.0$, inverse temperature $\beta=0.05$)}
                \label{fig:CiVt_wm}
            \end{figure}

            In this section we will go beyond the computation of the global reaction-time $\overline{t}_R$ and analyze more carefully the full pathway reaction kinetics and, in particular, the effect of phase separation of the catalysts involved.
            We will consider the case in which this phase separation leads to condensates where the catalysts are homogenously distributed (not in patches), and study the dynamics in time of the substrate concentration, $C_{\mu}(t)$.
            In order to make an estimation for $C_{\mu}(t)$, we can assume that mass action kinetics holds in a WM-system and, thus, formulate the following set of ODEs:
            \begin{equation}
                \begin{aligned}
                \dot{C}_0 &= -k C_0\\[1ex]
                \dot{C}_\mu &= k (C_{\mu-1} - C_\mu)\ ,\ \forall \mu \geq 1
                \end{aligned}
                \label{eq:odes_wm}
            \end{equation}
            All reactions in Eq.~\ref{eq:cascade} are assumed to have the same reaction rate $k$ (congruent with the assumption that all the different kinds of catalysts occupy the same surface fraction).
            Solving these ODEs, with initial conditions corresponding to a given amount of $S_0$ introduced in a closed system, yields (for the WM-system):
            \begin{equation}
                C_{\mu}^{wm}(t) = k^{\mu}C_{0}(0)\frac{t^{\mu}}{\mu!}e^{-kt}
                \label{eq:meanf_wm}
            \end{equation}
            with $C_0(0)$ being the initial amount of $S_0$ on the lattice.
            We have tested these analytical expressions against full microscopic simulations. 
            Our results are shown in Fig.~\ref{fig:CiVt_wm}: the  time traces $C_\mu(t)$ from the numerical simulation are reproduced quantitatively using Eq.~\ref{eq:meanf_wm} by just fitting one free parameter.
            
            This result is to be expected, since it is known that the law of mass action emerges statistically in a well mixed microscopic framework, like the one of our simplified model.
            Based on our previous simulation results we can propose here a simple extension of mass action kinetics in order to study PS systems.
            For this we modify the ODEs of Eq.~\ref{eq:odes_wm} by assuming that the reaction rates $k$ are all the same, except for the first reaction ($k_0$):
            \begin{equation}
                \begin{aligned}
                \dot{C}_0 &= -k_0 C_0\\[1ex]
                \dot{C}_1 &= k_0 C_0 - k C_1\\[1ex]
                \dot{C}_\mu &= k (C_{\mu-1} - C_\mu)\ ,\ \forall \mu \geq 2
                \end{aligned}
                \label{eq:odes_ps}
            \end{equation}
            This is to take into account that the first step is dominated by the time it takes the initial substrate to find the condensate.
            Solving these ODEs yields for a (homogeneous) PS-system:
            \begin{equation}
                \begin{aligned}
                C_{0}^{ps}(t) &= C_{0}(0)e^{-k_0t}\\[1ex]
                C_{\mu}^{ps}(t) &= k_{0}k^{\mu-1}C_{0}(0)e^{-kt}\cdot f_\mu ,\quad \mu\geq 1 \\[1ex]
                f_\mu &= \frac{ e^{(k-k_0)t} - \sum_{m=0}^{\mu-1} \frac{ ((k-k_0)t)^m}{m !} }{(k-k_0)^{\mu}}
                \end{aligned}
                \label{eq:meanf_ps_app}
            \end{equation}
            \begin{figure}[ht]
                \centering
                \includegraphics[width=0.8\textwidth]{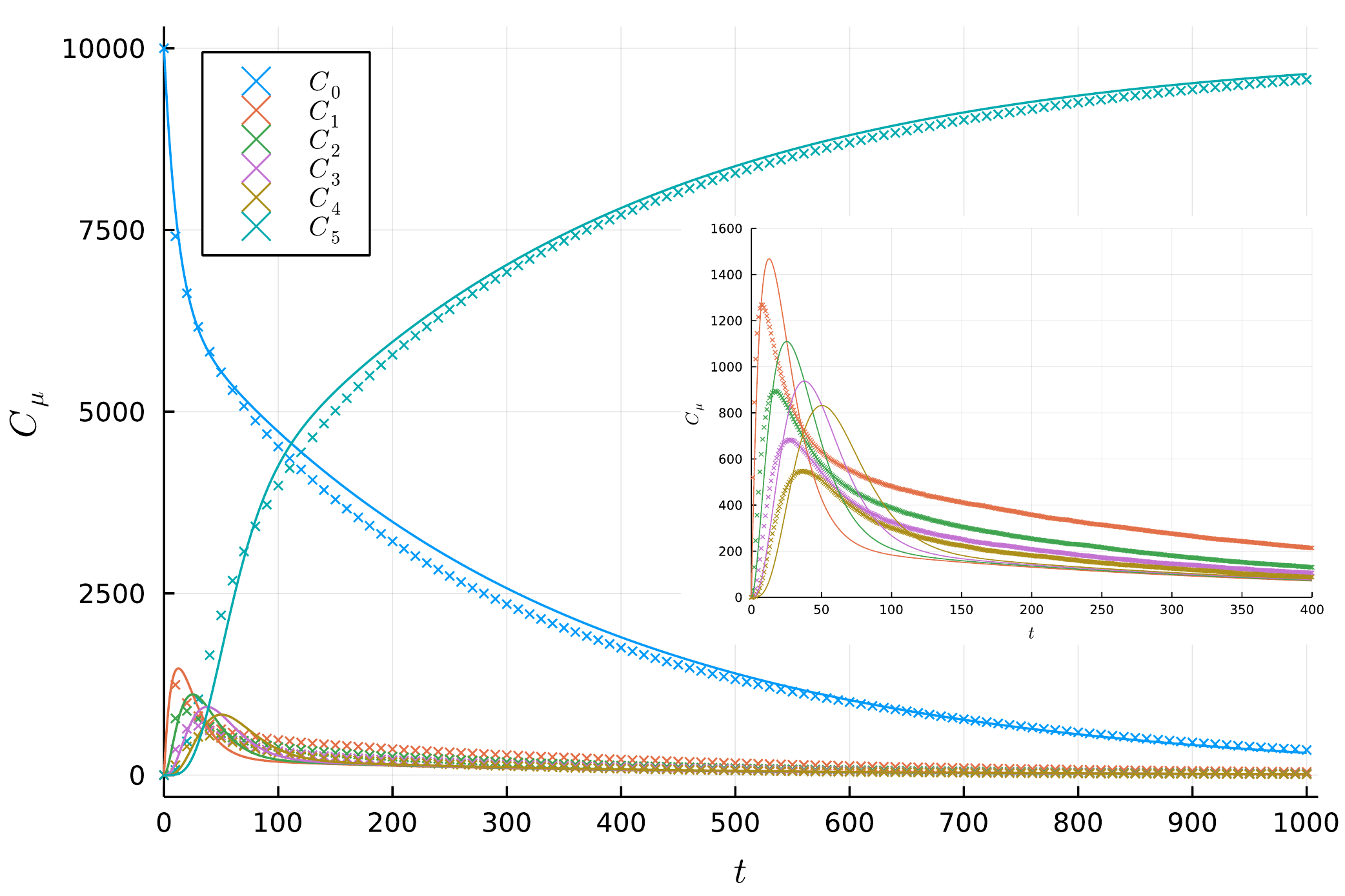}
                \caption{Time dynamics for the substrate concentrations $C_\mu(t)$ from full microscopic simulations (points) and the analytical solution of the (modified/adapted) mass action ODEs (lines) in a PS-system (lattice size $d=100$, pathway length $L=5$, catalyst surface fraction $\phi_{tot}=0.3$, catalyst-catalyst interaction strength  $J=1.0$, substrate-catalyst interaction strength $I=1.0$, inverse temperature $\beta=5$). Inset: intermediate concentrations.}
                \label{fig:CiVtps}
            \end{figure}
            Again, we tested our approach against full microscopic simulations and the results are shown in Fig.~\ref{fig:CiVtps}.
            Although a relatively good agreement was found for the trajectories of the initial substrate and final product, the trajectories of the intermediates are not so precisely reproduced (see the inset of Fig.~\ref{fig:CiVtps}).
            In any case, our approach of modifying mass action kinetics by taking into account the difference in the initial reaction step captures qualitatively the system behavior.
            The formation of the condensate creates a spatial asymmetry (or scale effect) that is transduced, so to speak, to a temporal asymmetry. 
            When the process of phase separation leads to a `homogeneous' condensate, this would take place just once.
            When the condensate is ``patchy'' (see previous section) one could project, from this first result, that subsequent `scale-effect transduction-steps' could occur in the system.
            Further research is required in this direction.
    
        \subsubsection{Phase separation process suppresses feedback-induced oscillations}
        \label{sec:res_thry2}
            In this final subsection of results we consider a more realistic and complex setting for our reaction pathway, having in mind a biological, cellular context.
            With that aim, more specifically: $(i)$ we `open' the system, allowing the in-flow of the initial substrate, $S_0$, as well as the out-flow of the final product, $P$; then $(ii)$ we introduce a `product-inhibition' feedback mechanism, by which $P$ hinders the inflow of $S_0$.   
            This is actually a common homeostatic mechanism, operating in many linear metabolic pathways (typically acting through the allosteric regulation of the first enzyme by the final product) \cite{heinrich2012regulation}.
            The new conditions can be summarized by the following picture:
            % \vspace{3ex}
            % \begin{equation}
            %     \ce{ {\ccoord{a}} -> S_{0} ->[E_{1}] S_{1} ->[E_{2}] S_{2} \quad \dots \quad S_{L-1} ->[E_{L}] {\ccoord{b}} P -> }
            %     \label{eq:cascade2}
            % \end{equation}
            % \tikz[overlay,remember picture]{
            %     \draw[-|] (b) -- ++(0,0.7em) -| (a)
            %     node[below, near start] {};
            % }
            \begin{equation}
                \includegraphics[valign=c]{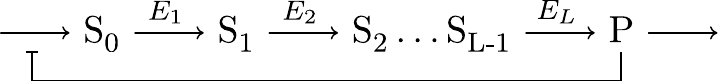}
                \label{eq:cascade2}
            \end{equation}
            
            Given its importance in biochemistry and metabolism, such a reaction system has been extensively studied, using MAK, and it has been established that the feedback can induce  non-linear oscillations \citep{Hunding:1974}, where a necessary condition for the onset of oscillations is the homogeneity of reaction kinetic constants.
            
            Yet, our previous results show that the phase separation of the enzymes in a condensate would break the homogeneity of the kinetic steps, where the first step is dominated by the process in which the substrate has to reach the condensate. 
            Our framework thus predicts that the triggering of phase separation could cancel out oscillations that would be present in the WM-case.
            
            \begin{figure}[ht]
                \centering
                \begin{subfigure}[b]{0.48\textwidth}
                \includegraphics[width=\textwidth]{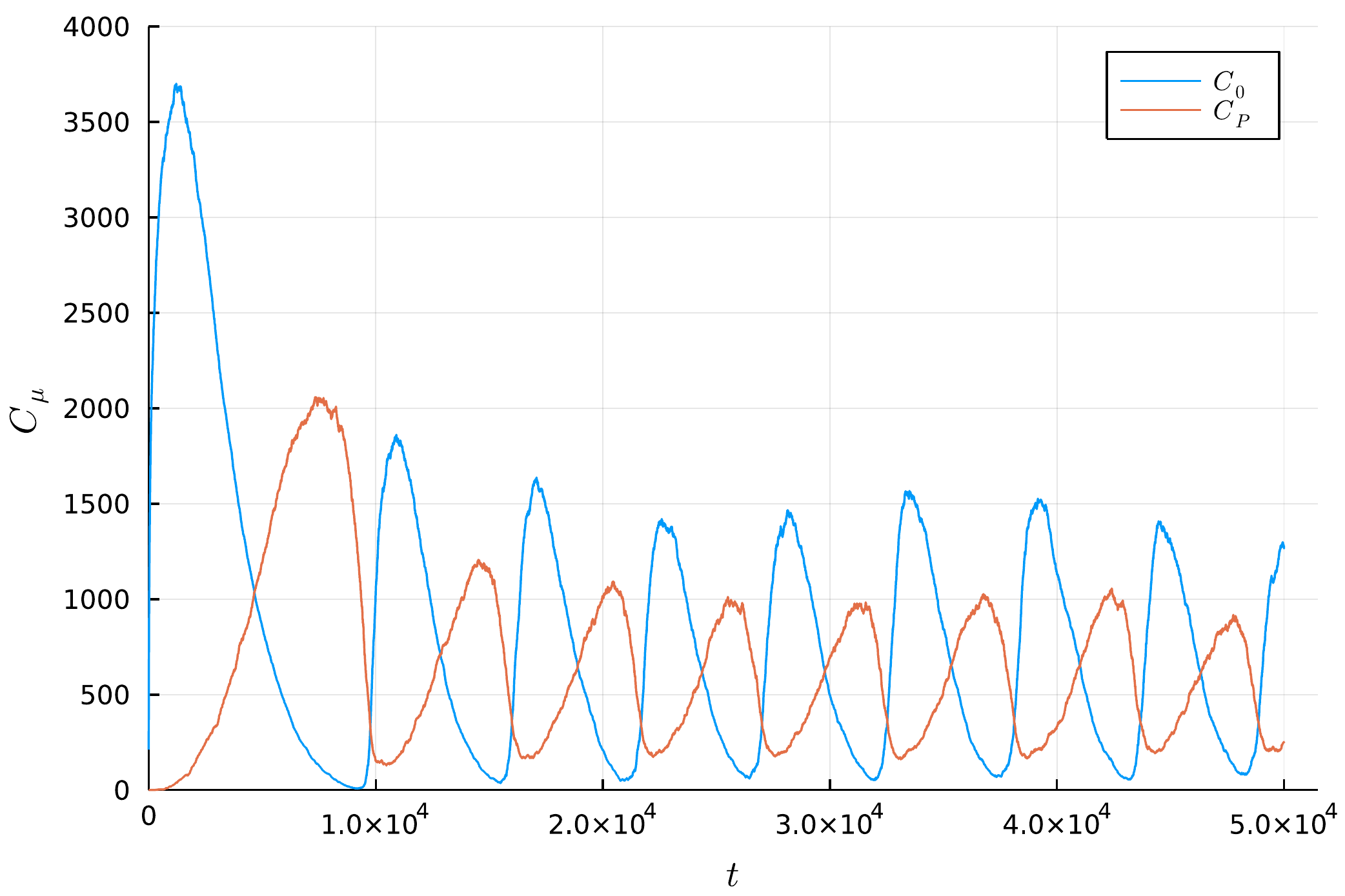}
                    \caption{}
                    \label{fig:FeedbInh_wm}
                \end{subfigure}
                \hspace{1ex}
                \begin{subfigure}[b]{0.48\textwidth}
                    \includegraphics[width=\textwidth]{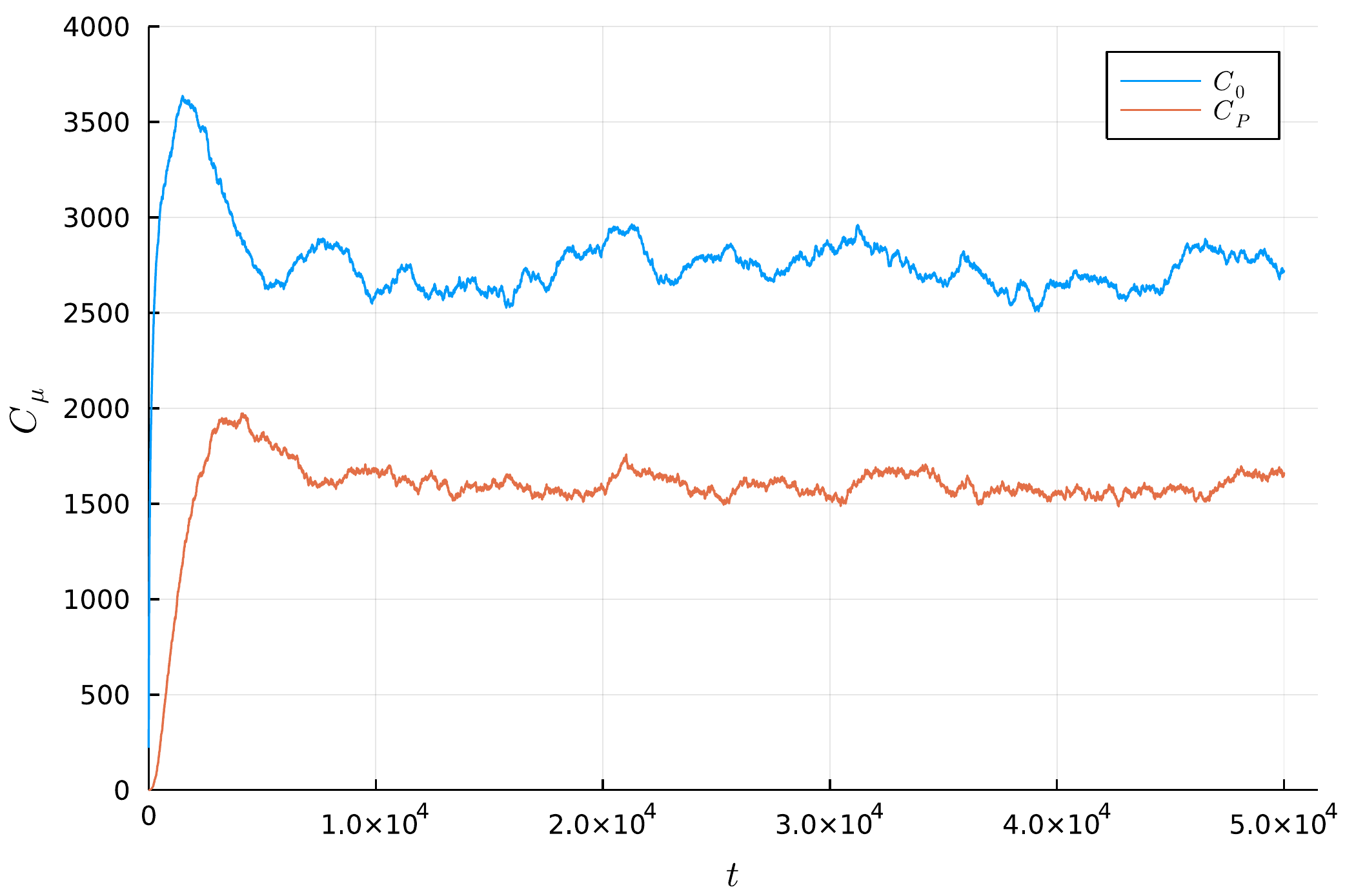}
                    \caption{}
                    \label{fig:FeedbInh_ps}
                \end{subfigure}
                \caption{Time trends of initial substrate and $C_0(t)$ and final product $C_P(t)$ concentrations  from microscopic  simulations (lattice  size $d=30$, pathway length $L=10$,  enzyme volume fraction $\phi_{tot}=0.3$ interaction strength $J=1.0$). \textbf{(a)} WM system: substrate-enzyme interaction $I=0.0$ and an inverse temperature $\beta=0.0$. \textbf{(b)} PS system: substrate-enzyme interaction $I=1.0$ and inverse temperature of $\beta=5$.}
                \label{fig:FeedbInh}
            \end{figure}
            
            Indeed, we tested this prediction against microscopic simulations and our results (shown in Fig.~\ref{fig:FeedbInh_wm}) confirm that oscillations in the concentrations of intial substrate $C_0(t)$ and final product $C_P(t)$ for the WM-system (Fig.~\ref{fig:FeedbInh_wm}) are effectively suppressed in a PS-system (Fig.~\ref{fig:FeedbInh_ps}).
            Interestingly, many pathways in central metabolism present a similar product inhibition motif, supposedly leading to oscillatory behaviour, but the latter is rarely observed -- with the notable exception of glycolysis \citep{sel1968self,madsen2005mechanisms}. Our results are in agreement with this fact, and could be taken as a first theoretical hint about where to search, although a proper explanation for it should include both (i) a more suitable modelling of metabolic processes and (ii) a more comprehensive account of the spatial constraints involved (not only LLPS-related ones).

    \section{Conclusion and outlook:}
      
        The analytical calculations and computer simulations performed, according to our modelling assumptions, suggested that when liquid-liquid phase separation drives a group of catalysts into a condensate there are, indeed, important effects on their collective action over a linear, multi-step reaction pathway.
        A first important observation is that completion of the reaction pathway benefits from phase separation if there is a minimal affinity between catalysts and substrates, so that the condensate effectively becomes a micro-environment for the multi-step reaction to proceed.
        The mean time computed for the full pathway to yield a product molecule depends, quite naturally, on this parameter (the overall substrate-catalyst interaction strength), but also on several other factors.
        One is the length of the pathway: in the comparison between the WM- and PS-cases, we found that there is typically a critical length above which the overall process turns out to be faster under catalyst de-mixing.
        Thus, we can conclude that spatial organization via phase separation can make catalytic action more efficient when pathways are relatively long (in our simulation results, the lower bound was roughly five reaction steps, although this will depend on other parameters and constraints). 

        A less intuitive or less straightforward result (with important implications, we believe, for biological systems) is that such a critical pathway length changes with the dimension of the lattice: the bigger the system, the longer the critical length for which PS induces a significant difference in terms of accelerating the overall process (always relative to the WM-case, where that scaling effect is not present).
        Since all this is, of course, for a given volume-/surface-fraction occupied by the catalysts in the corresponding space, it introduces a limitation for this type of control mechanism.
        Therefore, it looks like there could be an optimum, intermediate range of sizes for PS to operate more readily (and, perhaps, also to be subject to regulation, alternating from one state to the other, back and forth -- although the latter is yet to be properly explored).
        Somehow, a compromise solution may have to be reached between the dimension of the condensates formed within the cytoplasm and the amount of different catalysts involved (namely, the length of the multi-step reactions to be kinetically controlled).

        In any case, this interesting size-effect will depend on the geometries and the various degrees of homogeneity/inhomogeneity present in the system.
        Within our theoretical treatment of the problem, strictly speaking, the condensate consists in a 2D-micro-environment, which may be suitable to capture the configuration of a set of catalysts coming together on a surface (e.g., on a lipid bilayer).
        Our results could be extended and generalized to 3D, but the scaling factor would accordingly change, and the required simulation work would become computationally  more demanding.
        It is also important to remark that the size-effect, which translates into a temporal delay, might `percolate' in the system, if the condensate is not homogeneous (i.e., if the catalysts brought together are not well-mixed, among them, and create further patterns, like the ``patches'' observed above, in section 3.1.3).
        Thus, a cascade or a multiple combination of effects of this kind could be operating at once.
        Asymmetries in the catalyst-catalyst interaction matrix (reflecting differences in their mutual affinities) could also lead to diverse scenarios, to be addressed and more carefully analysed in future work.
        But our current approach already illustrates that the space of dynamic possibilities for a real cell is very rich, and varies according to its physiological conditions and the 2D or 3D configurations that its components may fall into.
        
        The extent to which previous MAK-based models should be modified to account for this complexity will depend on the phenomenon under exploration and level of refinement to be accomplished.
        As we showed, a rough qualitative approximation should take care of the initial reaction time involved whenever a condensate is formed.
        But this amendment may have to be repeated several times, at each scale in which PS effectively acts as a driving force to generate condensates.
        The most promising avenue of investigation, however, will be to ``open'' the system and analyse what should happen in non-equilibrium conditions, like we do for a relatively simple case in the last part of our work.
        Diverse phenomena that are in principle expected to occur under WM-conditions, like oscillatory behaviour in the presence of a feedback inhibition mechanism, might get cancelled -- as we demonstrated, and is in good agreement with metabolic phenomenology.
        Still, many other correlations and multiple cross-effects are bound to emerge, complicating the picture.
        For instance, even staying within our 2D framework, an intriguing possibility would be to consider that the initial substrate and the final product of the reaction chain are ``solvent molecules'' (e.g. lipids), whose balance would actually modify the properties of the physical environment (that is, the lattice itself) where PS processes take place, an aspect that we leave for future developments.

\section*{Acknowledgements}
    This work has been supported by the Horizon 2020 Marie Curie ITN (“ProtoMet”—Grant Agreement no. 813873 with the European Commission), within which NL obtained a PhD fellowship. 
    Both NL and KR-M acknowledge support from the Basque Government (IT1668-22), the Spanish Ministry of Science and Innovation (PID2019-104576GB-I00) and the John Templeton Foundation (grant 62220).
    OT thanks the Biofisika Institute for kind hospitality during the development of this work.
    OT acknowledges the Faculty of Mathematics and Physics of the Charles University (Prague, Czech Republic) where he is enrolled as a PhD student.
    DDM and KN thank the FBB (Fundación Biofísica Bizkaia) for support.
	
	%\noindent\textbf{Conflict of Interest}\par
	%\noindent The authors declare that there is no conflict of interest.

    %====Bibliography======================================================

	\bibliography{refs.bib}
	%======================================================================
    
    % \renewcommand{\theequation}{A.\arabic{equation}}
     % \appendix

    \newpage
    \section*{Appendix}

    \renewcommand{\thefigure}{A.\arabic{figure}}
    \setcounter{figure}{0}
    \renewcommand{\theequation}{A.\arabic{equation}}
    \setcounter{equation}{0}
    \renewcommand{\thesection}{A}
    \setcounter{section}{0}
    \section{Estimating the reaction time}
    \label{sec:tR}
        
        In general we have a lattice that contains $L$ different catalysts, i.e. the length of the reaction-cascade is $L$, the total volume/surface fraction of all catalysts is $\phi_{tot}$ and we assume that each type of catalyst occupies the same volume/surface fraction, thus $\phi_\mu$ the volume/surface fraction of a certain catalyst type $E_{\mu}$ is $\phi_\mu= \phi_{tot}/L$.
        We want to make an estimation for $t_R$, the time it takes to finish the reaction-cascade.
        In the context of the microscopic simulation this corresponds to the time it takes for a random walk on the lattice to encounter all $L$ catalysts in the correct order.
        There are two cases: the well mixed system and the phase-separated system, which we will analyze in the following.
                
        \subsection{Well-mixed system}
        \label{sec:tR_wm}
                    
            In a well mixed system the catalysts are homogeneously distributed in the lattice. The probability $p_{\mu}$ for a substrate particle $S_{\mu-1}$ to randomly jump to a lattice site, occupied by a catalyst $E_{\mu}$ is proportional to its concentration and, thus, to its volume/surface fraction $\phi_{\mu}$, i.e. $p_{\mu}\sim x_{\mu}$. The mean time for a single particle to reach a catalyst $E_{\mu}$ is then 
            
            \[
                \overline{t}_{\mu} = \frac{1}{p_{\mu}}\sim\frac{1}{\phi_{\mu}} = \frac{L}{\phi_{tot}}
            \]
    
            This is in principle the mean time for an individual reaction to finish. We can assume that the mean reaction-time $\overline{t}_R$ is proportional to the sum of all the induividual reaction times, i.e.
            
            \[
                \overline{t}_R\sim\sum_{\mu}^{L}t_{\mu}=\sum_{\mu}^{L}\frac{L}{\phi_{tot}}=L\frac{L}{\phi_{tot}}
            \]
            
            We thus have for $\overline{t}_R^{\ wm}$, the mean reaction-time in a well-mixed system:
            
            \begin{equation}
                \boxed{
                \overline{t}_R^{\ wm}=a\frac{L^2}{\phi_{tot}}
                }
                \label{eq:wm_time}
            \end{equation}
        
            with $a$ being a proportionality constant.
    
        \subsection{Phase-separated system}
        \label{sec:tR_ps}
                
            In a phase separated system, the catalysts will eventually start forming a droplet. Thus the mean-reaction time, i.e. the time to find the catalysts in the correct order is delayed by the time to encounter the droplet. We can thus assume that
            
            \begin{equation}
                \overline{t}_R = \overline{t}_{R}^{\ b} + \overline{t}_{h}
                \label{eq:timesplit}
            \end{equation}
            
            with $t_{R}^{\ b}$ the reaction time within the droplet and $\overline{t}_{h}$ the time to reach/hit the droplet (i.e the `hitting-time'). In other words, we effectively split the problem into two parts.

            \begin{figure}[ht]
                \centering
                \includegraphics[width=0.25\linewidth]{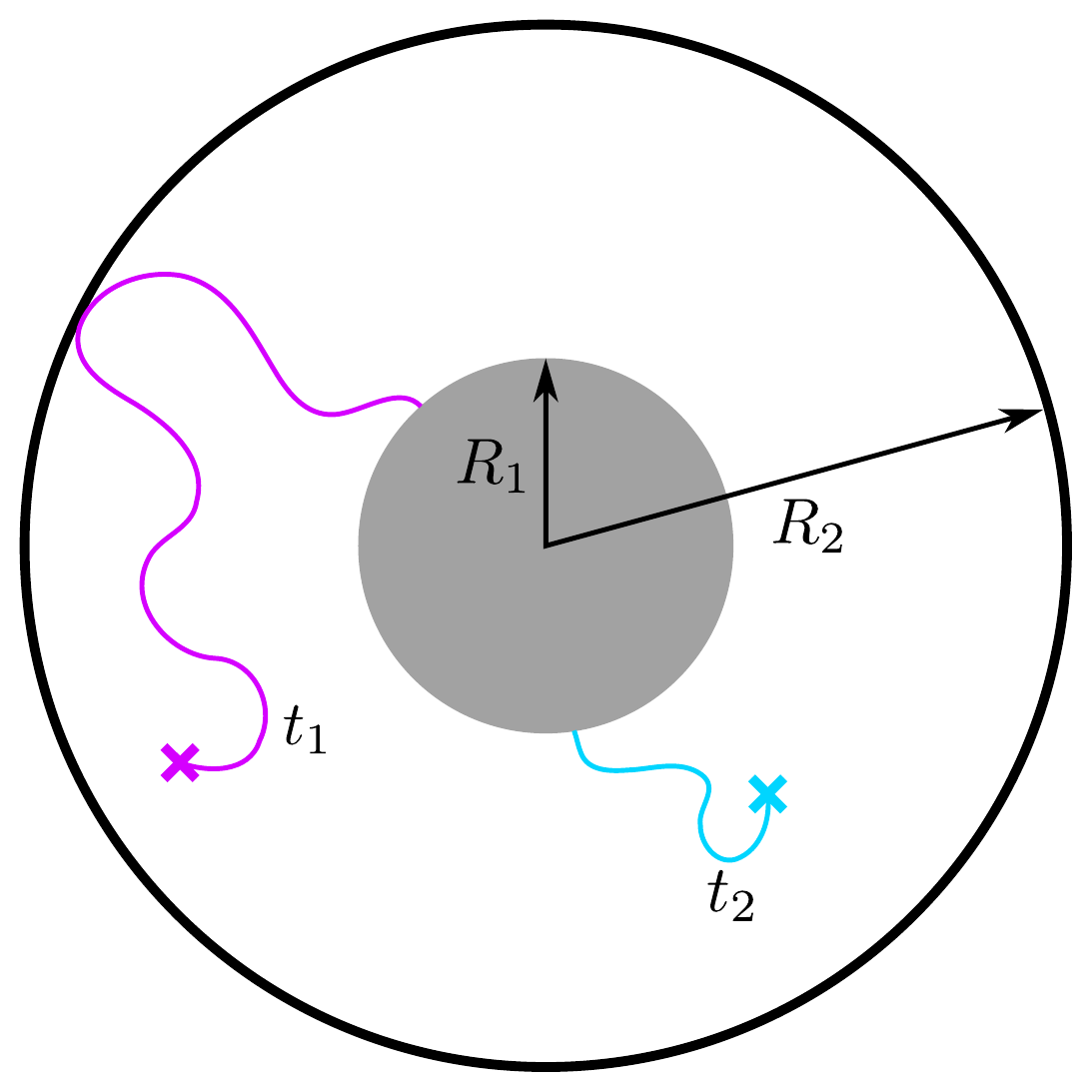}
                \caption{To get the `hitting-time' $\overline{t}_{h}$ we assume a spherical system with a reflective boundary-condition at $R_2$ and absorbing boundary-condition at $R_1$}
                \label{}
            \end{figure}
            
            For the reaction time inside the droplet we can make a simple assumption: as the particles stay inside the droplet and the catalysts are homogeneously distributed inside the droplet we can regard this situation as a well-mixed system with a total concentration of $\phi'_{tot}\sim 1$, thus from Eq.(\ref{eq:wm_time}) we have:
            
            \begin{equation}
                \overline{t}_{R}^{\ b} = aL^2
                \label{eq:blb_time}
            \end{equation}
            
            The mean time for single particle to reach the droplet $T(\vec{x})$ is a function of the position $\vec{x}$ and it verifies the Poisson Equation:
    	
    		\[
    		\nabla^2 T(\vec{x}) = -\frac{1}{2D}
    		\]
    		
    		In spherical coordinates and with $T(r)$ only depending on the radial-coordinate, this reduces to a second-order inhomogeneous ODE:
    		
    		\[
    		\begin{aligned}
    			\frac{1}{r}\frac{\partial}{\partial r}\left(r\frac{\partial}{\partial r }T\right) + \frac{1}{2D} &= 0\\[1ex]
    			T^{''}+\frac{1}{r}T^{'}+\frac{1}{2D} &= 0
    		\end{aligned}
    		\]
            
            In general, the solution for an inhomogeneous ODE is the sum of the solution for the homogeneous ODE and  particular solution, i.e.
            
            \[
                T = T_h + T_p
            \]
            
            For the homogeneous ODE we have:
            
            \[
                T_h^{''}+\frac{1}{r}T_h^{'} = 0
            \]
            
            Using a substitution method ($T_h^{'}=u$) to reduce this to a first order ODE we can get for the homogeneous solution:
            
            \[
                T_h = A\log r+B
            \]

            Through standard techniques (polynomial ansatz) we get for the particular solution:
            
            \[
                T_p = -\frac{1}{8D}r^2+C
            \]
            
            Thus collecting $T_h$ and $T_p$ we have for T(r):
            
            \[
                T(r) = A\log r-\frac{r^2}{8D}+B
            \]
            
            For the constant factors $A,B$ we can use the following boundary conditions (BC) at $R_1,R_2$: $T(R_1)=0$ (absorbing BC) and $\nabla T(R_2)=0$ (reflecting BC). 

            Thus we finally have:
            
            \begin{equation}
            \begin{aligned}
                T(r) &= \frac{R_2^2}{4D}\log r-\frac{r^2}{8D}+\frac{R_1^2}{8D}-\frac{R_2^2}{4D}\log R_1\\
                &= \frac{R_2^2}{4D}\log\frac{r}{R_1}-\frac{1}{8D}\left(r^2+R_1^2\right)
                \label{eq:hittime1}
            \end{aligned}
            \end{equation}
            
            We now have to integrate over the whole volume
            
            \[
            \begin{aligned}
                \overline{t} &= \frac{1}{V}\int T(\vec{x})\text{d}\vec{x}\\
                &= \phi_{tot}\cdot 0 + \frac{(1-\phi_{tot})}{\pi R_2^2}\int_{R_1}^{R_2}T(r)2\pi r\text{d}r
            \end{aligned}
            \]
            
            where the first term corresponds to the probability to be inside the droplet, while the second term corresponds to the probability to be outside the droplet. 

            We thus have for the integral:
    
            \[
                \overline{t} =\frac{(1-\phi_{tot})R_2^2}{16D}\left[ -1+4\phi_{tot}-2\log \phi_{tot}-\phi_{tot}^2) \right]
            \]
    
            In the case of a regular lattice we can assume that $R_2\sim d$. This finally gives the hitting-time as:
            
            \begin{equation}
                \overline{t}_h(\phi_{tot},d)=\frac{d^2}{16D}(1-\phi_{tot})(4\phi_{tot}-1-2\log \phi_{tot}-\phi_{tot}^2)
                \label{eq:hittime2}
            \end{equation}

            Putting Eq.(\ref{eq:blb_time}) and Eq.(\ref{eq:hittime2}) together we get for $\overline{t}_R^{\ ps}$, the mean reaction time in a phase-separated system:
            
            \begin{equation}
                \boxed{\begin{aligned}
                    \overline{t}_R^{\ ps} &= aL^2 + \frac{d^2}{16D}(1-\phi_{tot})(4\phi_{tot}-1-2\log \phi_{tot}-\phi_{tot}^2)\\[1.5ex]
                    &= aL^2 + bd^2(1-\phi_{tot})g(\phi_{tot})
                \end{aligned}}
                \label{eq:ps_time}
            \end{equation}
    
\begin{figure}[ht]
                \centering
                \begin{subfigure}[b]{0.48\textwidth}
                \includegraphics[width=\textwidth]{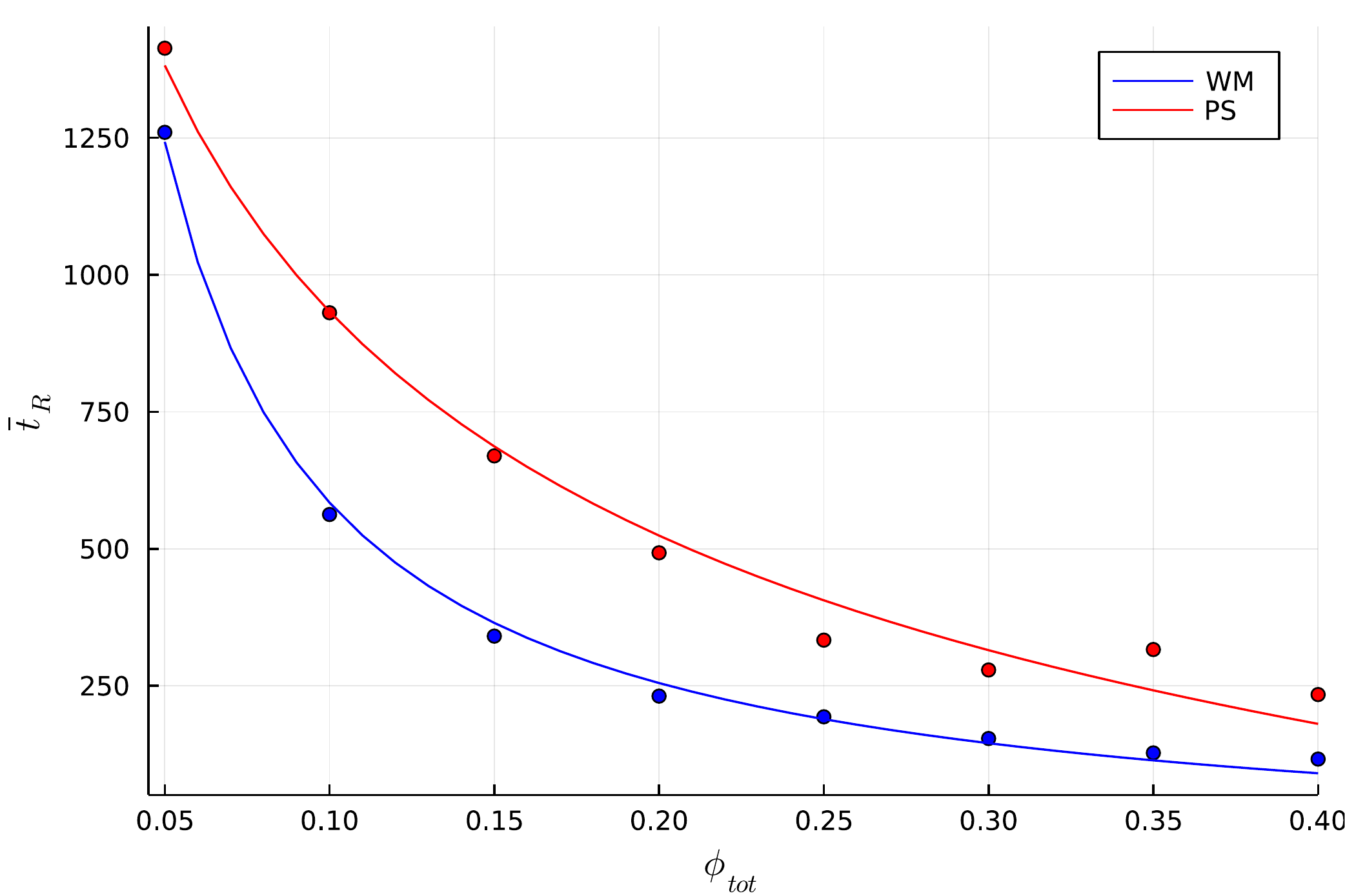}
                    \caption{}
                    \label{fig:tVphi5}
                \end{subfigure}
                \hspace{1ex}
                \begin{subfigure}[b]{0.48\textwidth}
                    \includegraphics[width=\textwidth]{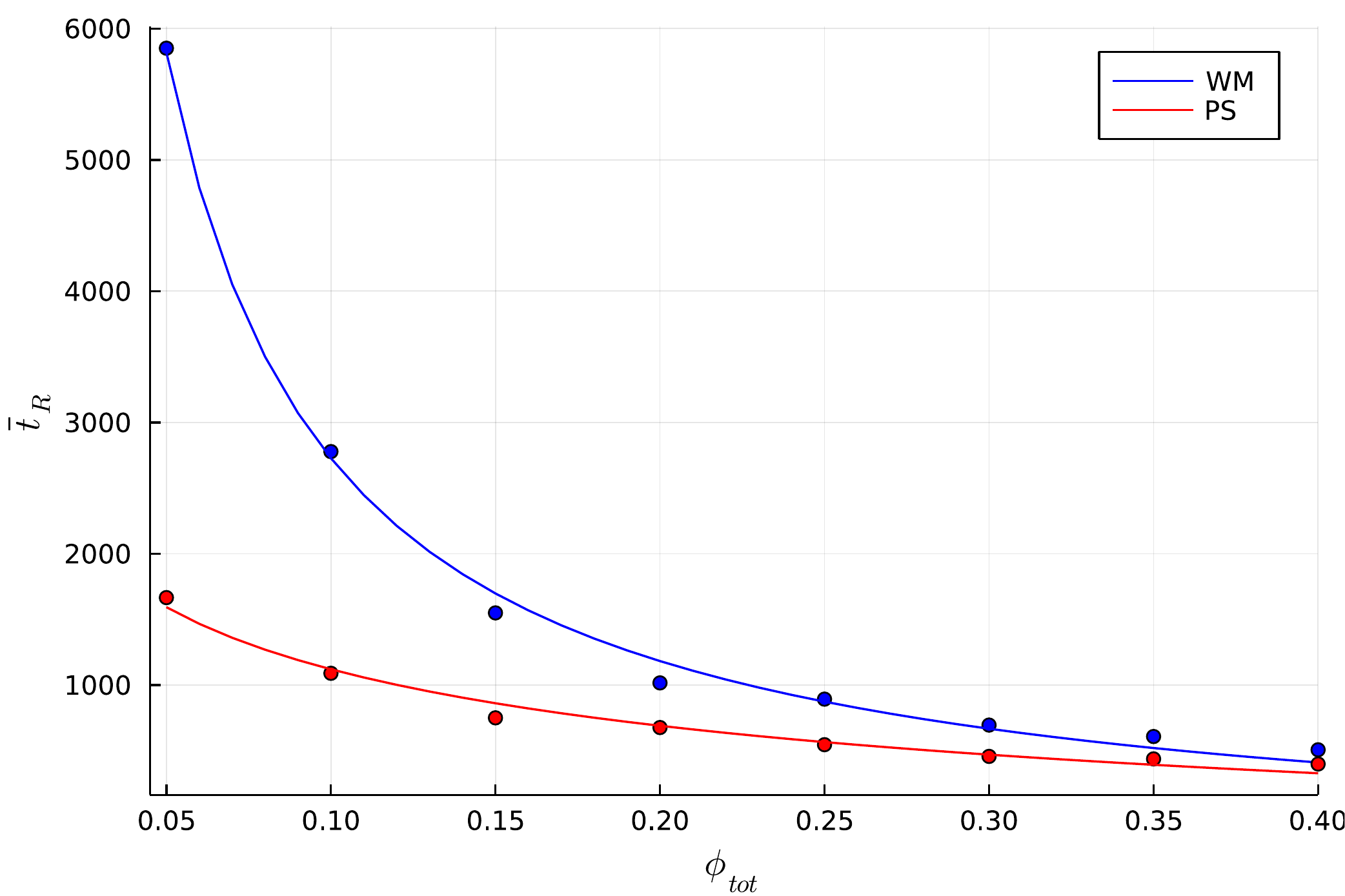}
                    \caption{}
                    \label{fig:tVphi10}
                \end{subfigure}
                \caption{Average reaction time $\overline{t}_R$ as a function of the total surface fraction $\phi_{tot}$ ($J=1.0,d=50$). Simulations were carried out for \textbf{(a)} $L=5$ and \textbf{(b)} $L=10$, comparing WM-systems (blue dots, $\beta=0.0$ and $I=0.0$) with PS-systems (red dots, $\beta=5.0$ and $I=1.0$). Solid lines:  fit of the simulation data using Eq.~\ref{eq:wm_time} and Eq.~\ref{eq:ps_time} for the WM- and the PS-systems, respectively.}
                \label{fig:tVphi}
            \end{figure}

            Comparing Eq.(\ref{eq:wm_time}) and Eq.(\ref{eq:ps_time}) we see that $\overline{t}_R^{\ ps}$ is basically $\overline{t}_R^{\ wm}$ but with a different slope and an offset that is proportional to $d$ and $\phi_{tot}$.
            In the main text the dependence of the mean reaction time $\overline{t}_R$ on the pathway-length $L$, predicted by Eq.~\ref{eq:wm_time} and Eq.~\ref{eq:ps_time}, was tested (see Fig.\ref{fig:tVn}). In Fig.~\ref{fig:tVphi} we furthermore tested the dependence of $\overline{t}_R$ on the surface fraction $\phi_{tot}$, as predicted byt the same equations.

        \subsection{Distribution of reaction times}
        \label{sec:tR_dist}

            In the main manuscript as well as in the preceding sections, we have only discussed the average reaction time. The complete distribution of reaction times for all the substrate molecules is, in fact, significantly different across the two types of systems examined.

            \begin{figure}[ht]
                \centering
                \begin{subfigure}[b]{0.48\textwidth}
                \includegraphics[width=\textwidth]{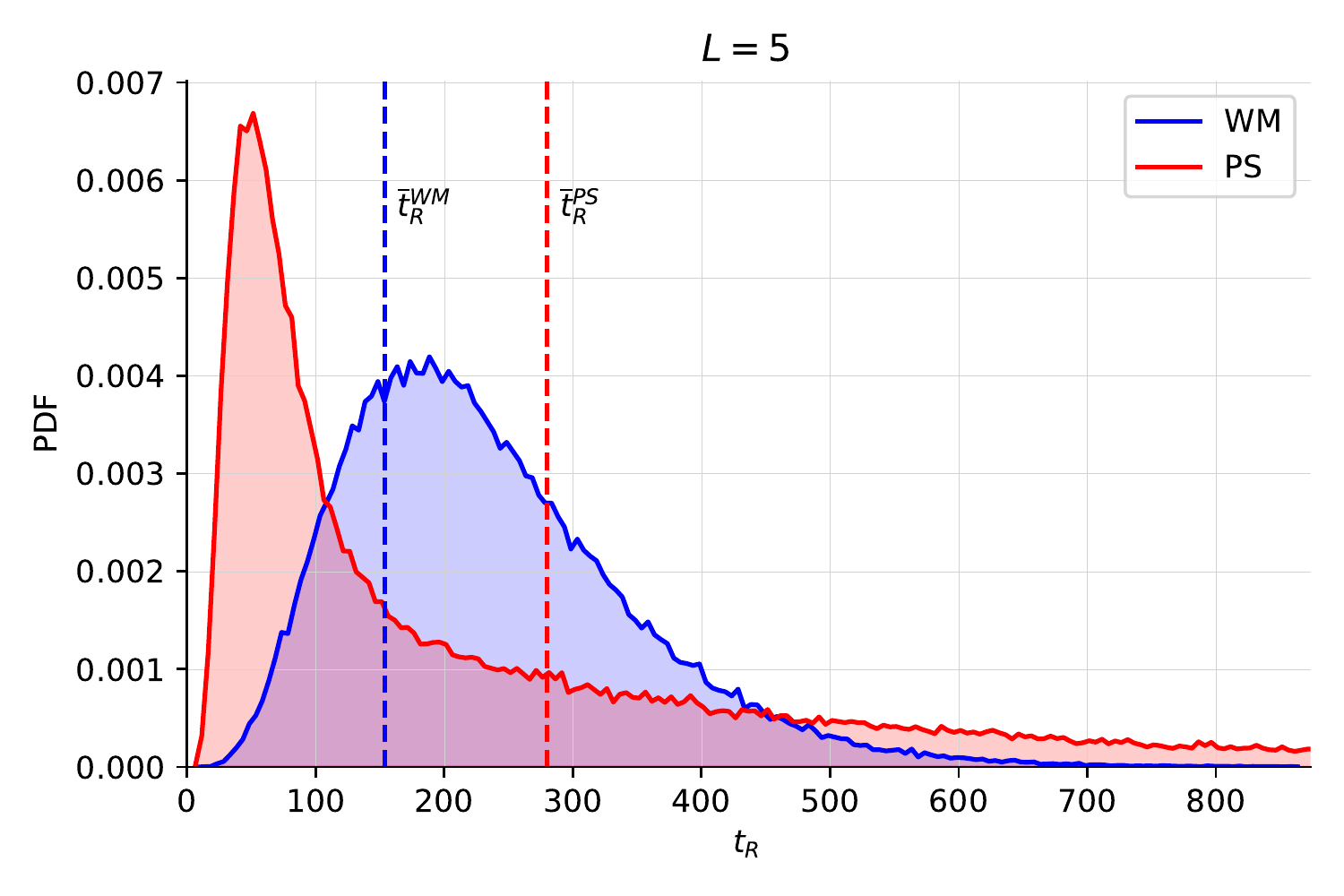}
                    \caption{}
                    \label{fig:meantimedist5}
                \end{subfigure}
                \hspace{1ex}
                \begin{subfigure}[b]{0.48\textwidth}
                    \includegraphics[width=\textwidth]{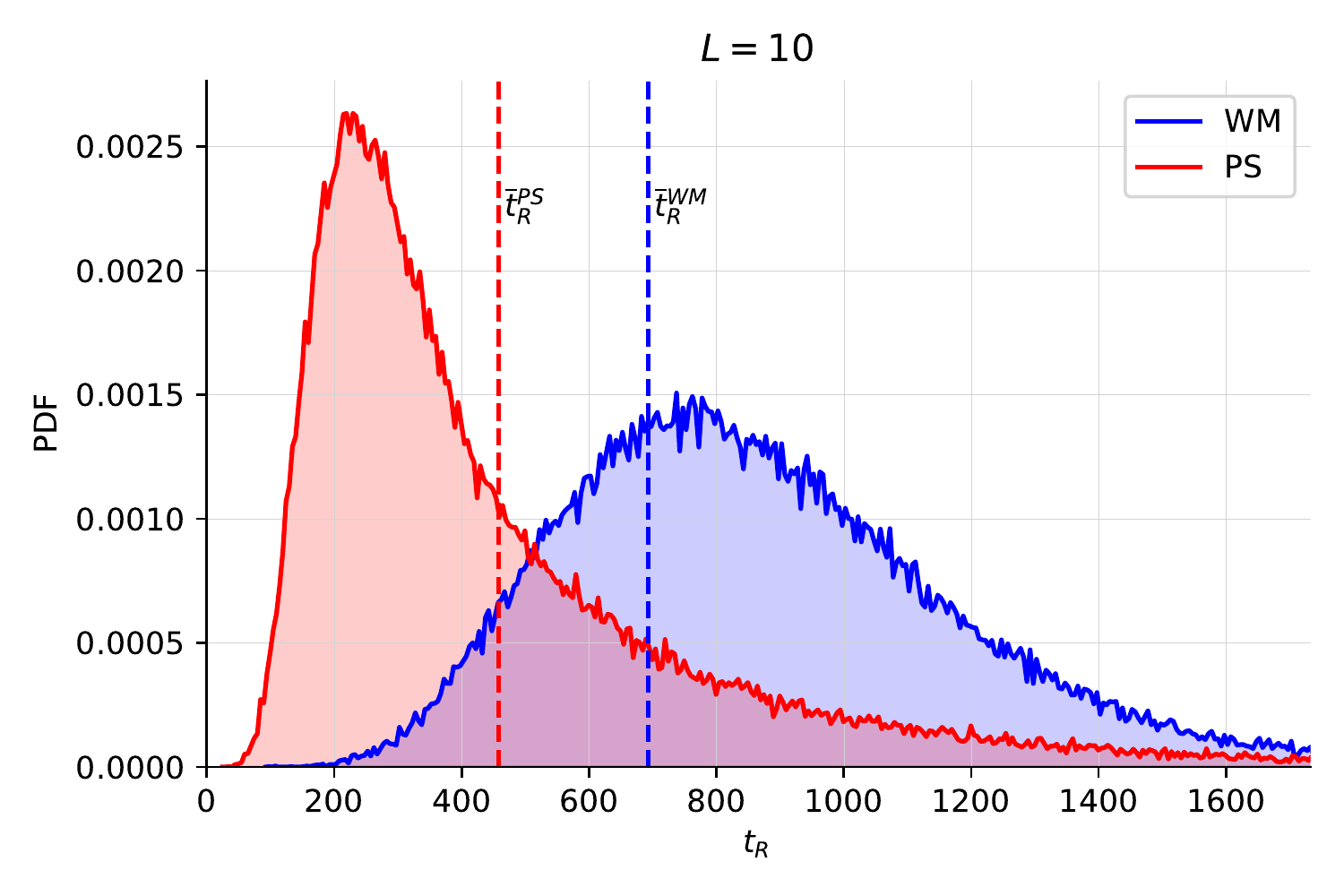}
                    \caption{}
                    \label{fig:meantimedist10}
                \end{subfigure}
                \caption{Distribution of total reaction times ($J=1.0,d=50$). Simulations were carried out for \textbf{(a)} $L=5$ and \textbf{(b)} $L=10$ for both WM-system (blue, $\beta=0.0$ and $I=0.0$) and PS-system (red, $\beta=5.0$ and $I=1.0$). Dashed lines show the average reaction time $\overline{t}_R$ for both cases.}
                \label{fig:meantimedist}
            \end{figure}

            As it is shown in Fig.~\ref{fig:meantimedist}, the distribution of reaction times in the PS-system exhibits a sharp initial peak and a long exponential tail, unlike the WM-system. Substrate molecules are initiated at random locations in the system during the simulations, and in case of the PS-system, some of them start from within the condensate ($\overline{t}_h=0$ in Eq.~\ref{eq:timesplit}) causing them to quickly complete all the reaction steps. This is the reason for the initial spike in the distribution of reaction times in the PS-system. The long exponential tail, on the other hand, comes from the distribution of hitting times $t_h$.

    \renewcommand{\thefigure}{B.\arabic{figure}}
    \setcounter{figure}{0}
    \renewcommand{\theequation}{B.\arabic{equation}}
    \setcounter{equation}{0}
    \renewcommand{\thesection}{B}
    \setcounter{section}{0}
    \section{Mean-field model for the change of concentration}
    \label{sec:mf}

        In principle our enzymatic cascade-reaction is an $L$-step sequential chemical reaction:

        \[
            \ce{S0 ->[k_0] S1 ->[k_1] S_2 \dots S_{L-1} ->[k_{L-1}] P}
        \]
        
        \vspace{1ex}
        
        We can formulate for these chemical reaction equations a set of coupled ODEs as:
        
        \begin{equation}
            \begin{aligned}
            	\dot{C}_0 &= -k_0C_0\\[1ex]
            	\dot{C}_{\mu} &= k_{\mu-1}\left[C_{\mu-1}\right] - k_{\mu}\left[C_{\mu}\right]\ ,\ \forall \mu\geq 1 \\[1ex]
            	\dot{C}_P &= k_{L-1}C_{L-1}
            \end{aligned}
            \label{eq:odes}
        \end{equation}
        
        In general we assume for the initial conditions: $C_0(0)=\text{const},C_{\mu}(0)=0\ \forall \mu \geq 1$ and $C_P(0)=0$. In the following we solve this set of ODEs for certain special cases.

        \subsection{Homogeneous reaction rate constants}
        \label{sec:mf_hom}
        
            We first want to look at the case of homogeneous reaction rate constants, i.e. $k_\mu=k, \forall \mu$.
            This corresponds to a well-mixed system.
            We get for Eq.(\ref{eq:odes}):
            
            \begin{equation}
                \begin{aligned}
                    \dot{C}_0 &= -kC_0\\[1ex]
                	\dot{C}_{\mu} &= k(C_{\mu-1} - C_{\mu})\ ,\ \forall \mu\geq 1 \\[1ex]
                	\dot{C}_P &= kC_{L-1}
                \label{eq:odes_wm_app}
                \end{aligned}
            \end{equation}
            
            The homogeneous ODE for $C_0(t)$ can be easily solved with the above established initial conditions:
            
            \[
                C_0(t) = C_0(0)e^{-kt}
            \]
            
            Turning now to the inhomogeneous ODEs for $C_{\mu}(t)$, we can use the method of the integrating-factor to get a recursive formula:
            
            \[
                C_{\mu}(t) = e^{-kt}\left[\int kC_{\mu-1}(t)e^{kt}\ \text{d}t + K\right]
            \]

           from which we get      
            \[
              C_{\mu}(t) = k^{\mu}C_0(0)\frac{t^{\mu}}{\mu!}e^{-kt}
            \]
            
            In order to solve the last ODE for $C_P(t)$, we can use mass conservation, to obtain:

            \begin{equation}
                \boxed{
                \begin{aligned}
                    C_{\mu}(t) &= k^{\mu}C_0(0)\frac{t^{\mu}}{\mu!}e^{-kt}\\
                    C_P(t) &= C_0(0)\left(1 - e^{-kt}\sum_{\mu=0}^{L-1} \frac{(kt)^{\mu}}{\mu!} \right)
                \end{aligned}
                }
                \label{eq:meanf_wm_app}
            \end{equation}

        \subsection{Homogeneous reaction rate constants with a different rate in first reaction}
        \label{sec:mf_inhom}
        
            A slight variation of the above system would be to leave the reaction rate constants the same for all reactions except the first one, i.e. $k_0, k_{\mu}=k, \forall \mu\geq 1$.
            This corresponds to a phase-separated system.
            We get for Eq.(\ref{eq:odes}):
            
            \begin{equation}
                \begin{aligned}
                    \dot{C}_0 &= -k_0C_0\\[1ex]
                    \dot{C}_1 &= k_0C_{0} - kC_1\\[1ex]
                	\dot{C}_{\mu} &= k(C_{\mu-1} - C_{\mu})\ ,\ \forall \mu\geq 2 \\[1ex]
                	\dot{C}_P &= kC_{L-1}
                \label{eq:odes_ps_app}
                \end{aligned}
            \end{equation}

            with the above established initial conditions.
            Thus,
            
            \[
                C_0(t) = C_0(0)e^{-k_0t}
            \]
            
         and we obtain in general, for $C_{\mu}(t)$ ($\mu\geq1$):

            \[
                C_{\mu}(t) = k_0k^{\mu-1}C_0(0)e^{-kt}\underbrace{\left[\frac{e^{(k-k_0)t}-\sum_{m=0}^{\mu-1}\frac{\left[(k-k_0)t\right]^m}{m!}}{(k-k_0)^{\mu}}\right]}_{=f_{\mu}}
            \]
            
            Finally, the last ODE for $C_P(t)$ can be integrated using, again, mass conservation, leading to:

            \begin{equation}
                \boxed{
                \begin{aligned}
                    C_0(t) &= C_0(0)e^{-k_0t}\\[2ex]
                    C_{\mu}(t) &= k_0k^{\mu-1}C_0(0)e^{-kt}\cdot f_{\mu}\ ,\quad \mu\geq1\\[2ex]
                    C_P(t) &= C_0(0)\left(1 - k_0e^{-kt}\sum_{\mu=0}^{L-1} k^{\mu-1}\cdot f_{\mu}\right)\\[2ex]
                    f_{\mu} &= \frac{e^{(k-k_0)t} - {\displaystyle\sum_{m=0}^{\mu-1}\frac{\left[(k-k_0)t\right]^m}{m!}}}{(k-k_0)^{\mu}}
                \end{aligned}
                }
                \label{eq:meanf_ps_app}
            \end{equation}

\end{document}